\documentclass[twoside,twocolumn,9pt]{article}
\usepackage{extsizes}
\usepackage[super,sort&compress,comma]{natbib} 
\usepackage[version=3]{mhchem}
\usepackage[left=1.5cm, right=1.5cm, top=1.785cm, bottom=2.0cm]{geometry}
\usepackage{balance}
\usepackage{mathptmx}
\usepackage{sectsty}
\usepackage{graphicx} 
\usepackage{lastpage}
\usepackage[format=plain,justification=justified,singlelinecheck=false,font={stretch=1.125,small,sf},labelfont=bf,labelsep=space]{caption}
\usepackage{float}
\usepackage{fancyhdr}
\usepackage{fnpos}
\usepackage[english]{babel}
\addto{\captionsenglish}{%
  
}
\usepackage{array}
\usepackage{droidsans}
\usepackage{charter}
\usepackage[T1]{fontenc}
\usepackage[usenames,dvipsnames]{xcolor}
\usepackage{setspace}
\usepackage[compact]{titlesec}
\usepackage{hyperref}

\newcommand{\CC}{\mathrm{CC}}
\newcommand{\R}{\mathrm{R}}

\newcommand{\etal}{\textit{et al.}}
\newcommand{\bra}[1]{\langle #1 \vert}
\newcommand{\ket}[1]{\vert #1 \rangle}
\newcommand{\bh}[2]{\mathbf{h}^{#1}\mathbf{D}^{#2}}
\newcommand{\bG}[2]{\mathbf{D}^{#1}\mathbf{G}(\mathbf{D}^{#2})}
\usepackage{epstopdf}

\definecolor{cream}{RGB}{222,217,201}
\DeclareMathOperator{\tr}{Tr}

\begin{document}

\pagestyle{fancy}
\thispagestyle{plain}
\fancypagestyle{plain}{
\renewcommand{\headrulewidth}{0pt}
}

\makeFNbottom
\makeatletter
\renewcommand\LARGE{\@setfontsize\LARGE{15pt}{17}}
\renewcommand\Large{\@setfontsize\Large{12pt}{14}}
\renewcommand\large{\@setfontsize\large{10pt}{12}}
\renewcommand\footnotesize{\@setfontsize\footnotesize{7pt}{10}}
\makeatother

\renewcommand{\thefootnote}{\fnsymbol{footnote}}
\renewcommand\footnoterule{\vspace*{1pt}%
\color{cream}\hrule width 3.5in height 0.4pt \color{black}\vspace*{5pt}} 
\setcounter{secnumdepth}{5}

\makeatletter 
\renewcommand\@biblabel[1]{#1}            
\renewcommand\@makefntext[1]%
{\noindent\makebox[0pt][r]{\@thefnmark\,}#1}
\makeatother 
\renewcommand{\figurename}{\small{Fig.}~}
\sectionfont{\sffamily\Large}
\subsectionfont{\normalsize}
\subsubsectionfont{\bf}
\setstretch{1.125} 
\setlength{\skip\footins}{0.8cm}
\setlength{\footnotesep}{0.25cm}
\setlength{\jot}{10pt}
\titlespacing*{\section}{0pt}{4pt}{4pt}
\titlespacing*{\subsection}{0pt}{15pt}{1pt}

\fancyfoot{}
\fancyfoot[LO,RE]{\vspace{-7.1pt}\includegraphics[height=9pt]{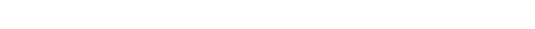}}
\fancyfoot[CO]{\vspace{-7.1pt}\hspace{11.9cm}\includegraphics{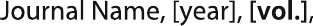}}
\fancyfoot[CE]{\vspace{-7.2pt}\hspace{-13.2cm}\includegraphics{RF}}
\fancyfoot[RO]{\footnotesize{\sffamily{1--\pageref{LastPage} ~\textbar  \hspace{2pt}\thepage}}}
\fancyfoot[LE]{\footnotesize{\sffamily{\thepage~\textbar\hspace{4.65cm} 1--\pageref{LastPage}}}}
\fancyhead{}
\renewcommand{\headrulewidth}{0pt} 
\renewcommand{\footrulewidth}{0pt}
\setlength{\arrayrulewidth}{1pt}
\setlength{\columnsep}{6.5mm}
\setlength\bibsep{1pt}

\makeatletter 
\newlength{\figrulesep} 
\setlength{\figrulesep}{0.5\textfloatsep} 

\newcommand{\topfigrule}{\vspace*{-1pt}%
\noindent{\color{cream}\rule[-\figrulesep]{\columnwidth}{1.5pt}} }

\newcommand{\botfigrule}{\vspace*{-2pt}%
\noindent{\color{cream}\rule[\figrulesep]{\columnwidth}{1.5pt}} }

\newcommand{\dblfigrule}{\vspace*{-1pt}%
\noindent{\color{cream}\rule[-\figrulesep]{\textwidth}{1.5pt}} }

\makeatother
%
\twocolumn[
  \begin{@twocolumnfalse}
{\includegraphics[height=30pt]{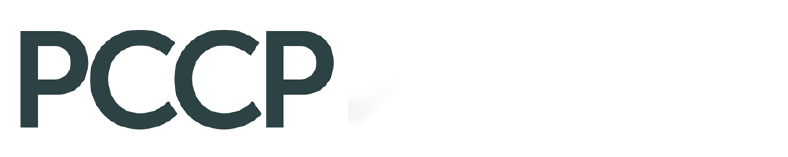}\hfill\raisebox{0pt}[0pt][0pt]{\includegraphics[height=55pt]{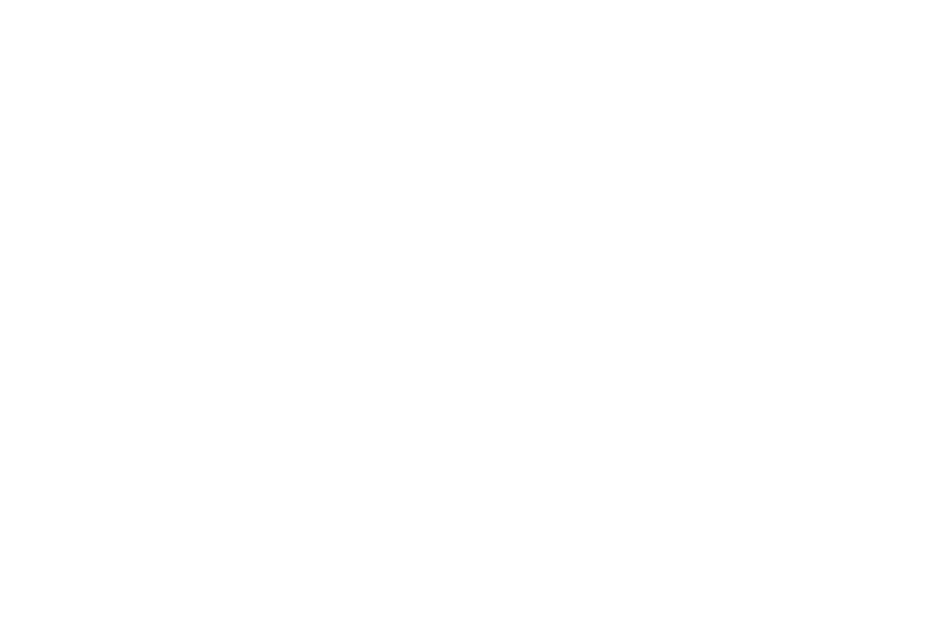}}\\[1ex]
\includegraphics[width=18.5cm]{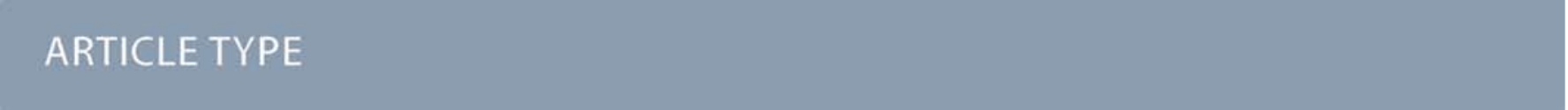}}\par
\vspace{1em}
\sffamily
\begin{tabular}{m{4.5cm} p{13.5cm} }
\includegraphics{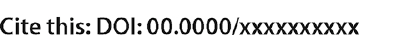} & \noindent\LARGE{\textbf{Combining Multilevel Hartree Fock and Multilevel Coupled Cluster with Molecular Mechanics: a Study of Electronic Excitations in Solutions
$^\dag$}} \\
\vspace{0.3cm} & \vspace{0.3cm} \\
 & \noindent\large{Linda Goletto,\textit{$^{a\ddag}$} Tommaso Giovannini,\textit{$^{a\ddag}$} Sarai D. Folkestad\textit{$^{a}$} and Henrik Koch\textit{$^{b}$}}$^{\ast}$ \\
\includegraphics{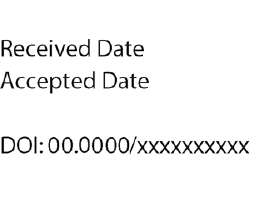} & \noindent\normalsize{We investigate the coupling of different quantum-embedding approaches with a third molecular-mechanics layer, which can be either polarizable or non-polarizable. In particular, such a coupling is discussed for the multilevel families of methods, in which the system is divided into an active and an inactive orbital spaces.
The computational cost of the resulting three layers approaches is reduced by treating the long-range interactions at the classical level.
The methods developed are tested against the calculation of the excitation energies of molecular systems in aqueous solution, for which an atomistic description of the environment is crucial to correctly reproduce the specific solute-solvent interactions, such as hydrogen bonding. In particular, we present the results obtained for three different moieties, acrolein, pyridine and para-nitroaniline, showing that an almost perfect agreement with experimental data can be achieved when the relevant physico-chemical interactions are included in the modeling of the condensed phase.} \\
\end{tabular}

 \end{@twocolumnfalse} \vspace{0.6cm}
]

\renewcommand*\rmdefault{bch}\normalfont\upshape
\rmfamily
\section*{}
\vspace{-1cm}


\footnotetext{\textit{$^{a}$~Department of Chemistry, Norwegian University of Science and Technology (NTNU), 7491 Trondheim, Norway}}
\footnotetext{\textit{$^{b}$~Scuola Normale Superiore, Piazza dei Cavalieri 7, I-56126 Pisa, Italy; E-mail: henrik.koch@sns.it }}
\footnotetext{\dag~Electronic Supplementary Information (ESI) available: data related to Figs. \ref{fig:acro_12_small_results_mlhf}-\ref{fig:acro_12_big_results}. See DOI: 10.1039/cXCP00000x/}
\footnotetext{\ddag~L.G. and T.G. contributed equally to this work.}

\section{Introduction}

The study of excited-state properties of systems has become one of the most important topics in computational chemistry. The theoretical investigation of such properties for molecules in the gas-phase can be performed by using different approaches, ranging from time dependent density functional theory (TD-DFT) to the most accurate equation-of-motion coupled cluster (EOM-CC) or coupled cluster response theory (CCRT).\cite{stanton1993equation,helgaker2014molecular,koch1990excitation,koch1990coupled,pedersen1997coupled} When a moiety is dissolved in a solution, a reliable description of the solute-solvent interactions is crucial to reproduce the experimental findings.\cite{cannelli2017understanding,carlotti2018evaluation,mennucci1998excited,jacquemin2011excited,corni2005electronic,cupellini2019electronic,mennucci2019multiscale} In fact, although the measured observable is due to the solute, it can be dramatically modified by the external environment, in particular when the solute is embedded in a strongly interacting solvent, such as water.\cite{giovannini2020csr,giovannini2020pccp}

Condensed systems are usually treated by resorting to the so-called focused approaches, where the attention is centered on the solute, which retains the final property, and the solvent is instead treated at a lower level of theory.\cite{warshel1976theoretical,warshel1972calculation,miertuvs1981electrostatic,tomasi1994molecular,orozco2000theoretical,tomasi2005,giovannini2020csr} The most widespread models have been developed in the context of quantum mechanical (QM)/classical approaches, in which the solute is 
described
at the QM level, whereas the environment is 
treated 
by means of classical mechanics, either at the continuum\cite{tomasi2005,tomasi1994molecular,mennucci2010continuum,mennucci2013modeling,lipparini2016perspective} or atomistic 
level
.\cite{senn2009qm,lin2007qm} The latter approaches overcome the well-known limits of the continuum description by retaining the atomistic description of the solvent, which is treated by means of Molecular Mechanics (MM) force field. Thus, the resulting QM/MM methods allow for the treatment of specific solute-solvent interactions, such as hydrogen bonding. In most QM/MM approaches, the QM and MM parts 
interact at the electrostatic level, including also the mutual polarization between the two regions in the polarizable QM/MM approaches.\cite{loco2016qm,curutchet2009electronic,olsen2011molecular,steindal2011excitation,boulanger2012solvent,jurinovich2014fenna,giovannini2019fqfmu,giovannini2019fqfmuder2,giovannini2019fqfmulinear} Therefore, the purely quantum forces, such as Pauli repulsion and dispersion, are usually neglected, although they may play a significant role.\cite{giovannini2017disrep,giovannini2019eprdisrep,giovannini2019quantum,gokcan2018qm,Gordon_JPCA_EFP,curutchet2018density} 
Moreover, the accuracy of QM/MM methods strongly depends on the quality of the parametrization of the classical force field. In order to overcome such limitations, quantum embedding approaches can be used.\cite{gordon2013accurate,gordon2007effective,sun2016quantum,knizia2013density,chulhai2018projection,chulhai2017improved,wen2019absolutely,ding2017embedded,goodpaster2012density,goodpaster2014accurate,manby2012simple,goodpaster2010exact,zhang2020multi,ramos2015performance,pavanello2011modelling} The latter are usually based on 
partitioning 
the density of the investigated system into two parts
: an active, which leads the property, and an inactive part, which perturbs the active moiety.  

Among the several quantum-embedding methods, a variety of multilevel (ML) approaches based on Hartree Fock (HF) and coupled cluster (CC) theories have been developed in the last years.\cite{myhre2013extended, myhre2014multi,myhre2016multilevel,saether2017density,hoyvik2020convergence,giovannini2020mlhf_ab,folkestad2019multilevel,folkestad2020equation,folkestad2020multilevel} These MLHF and MLCC methods rely on a partitioning of the orbitals into active and inactive orbital sets. 
In MLHF, the active occupied molecular orbitals (MOs) are obtained by means of a Cholesky decomposition of some initial density of the system.\cite{aquilante2003theoretical, sanchez2010cholesky}  
Active virtual orbitals can be obtained from orthonormalized projected atomic orbitals (PAOs).\cite{pulay.1983,saebo1993} 
The active and inactive MOs can also be localized in their pre-defined spatial regions using the MLHF-AB method, which allows for a further reduction in the computational cost.\cite{giovannini2020mlhf_ab} 
The MLHF(-AB) wave function can be used as a reference state for following CC calculations (CC-in-MLHF). 

In MLCC, the higher order excitation operators included in the cluster operator are restricted to a set of active MOs.\citep{myhre2014multi} This active orbital space can, for instance, be selected as in MLHF---using Cholesky orbitals and PAOs---or by using correlated natural transition orbitals (CNTOs).\cite{hoyvik2017correlated} 
In the CC-in-HF family of methods,\cite{sanchez2010cholesky,eT2020,folkestad2020multilevel} the cluster operator is fully restricted to the active orbital space; the active MOs are selected using Cholesky orbitals and PAOs.
The reference state is represented by the full HF wave function, and therefore the CC-in-HF method is more accurate than a CC calculation using an MLHF reference (CC-in-MLHF), whose accuracy is intrinsically determined by the quality of the decomposed density matrix.

In this work, we assess the accuracy of all the aforementioned approaches to the reproduction of the vacuo-to-water solvatochromic shifts.\cite{reichardt1992solvatochromism,reichardt1994solvatochromic} In order to effectively take into account the long range interactions, which are dominated by the electrostatic contribution, we propose different three-layer approaches which couple the CC-in-MLHF, CC-in-HF or MLCC with a MM region. In particular, the water molecules closest to the molecules are included in the QM region, whereas the remaining waters are treated by means of the polarizable fluctuating charge (FQ) force field,\cite{cappelli2016integrated,giovannini2020csr,giovannini2020pccp} which has been proved to be particularly suitable for describing the water medium.\cite{giovannini2020csr} 
Within the proposed schemes, the electrostatic, polarization, Pauli repulsion and dispersion interactions between the solute and the closest water molecules are taken into account, together with the long-range electrostatic and polarization contributions. 

The manuscript is organized as follows. In the next section, we outline the computational methods. In particular, we focus on CC2-in-HF, CC2-in-MLHF(-AB) and on MLCC2, and we present their coupling with a third MM layer, which can either be non-polarizable or polarizable (QM/FQ).
We then discuss the computational protocol applied in the calculations.
The developed methods are applied to the prediction of the vacuo-to-water solvatochromic shifts of acrolein (ACRO), para-nitroaniline (PNA) and pyridine (PY). A summary and the future perspectives of the approaches conclude the manuscript.

\section{Theoretical Models}

In this section, we briefly recall the different theoretical methods that are used to simulate the excitation energies of molecular systems in solution. In particular, we present the theory for non-polarizable QM/MM and polarizable QM/FQ, MLHF, MLHF-AB, CC2, MLCC2 and CC2-in-HF. Also, the coupling between the QM methods and the MM region 
is discussed.


\subsection{Non-polarizable QM/MM and polarizable QM/FQ}

In QM/MM calculations the studied system is partitioned into two parts, which are treated at the QM and MM level, respectively. The energy of the system can be written as:
\begin{equation}
    E = E_{\text{QM}} + E_{\text{MM}} + E^{\text{int}}_{\text{QM/MM}}\ ,
\end{equation}
where, $E_{QM}$, $E_{MM}$ and $E^{\text{int}}_{\text{QM/MM}}$ are the QM, MM and their interaction energies, respectively. The different QM/MM approaches developed in the literature mainly differ 
in
the description of the interaction energy $E^{\text{int}}_{\text{QM/MM}}$. In most QM/MM models, $E^{\text{int}}_{\text{QM/MM}}$ reads: 
\begin{equation}
E^{\text{int}}_{\text{QM/MM}} = E^{\text{ele}}_{\text{QM/MM}} + E^{\text{pol}}_{\text{QM/MM}} \ ,
\label{eq:en_int}
\end{equation}
where $E^{\text{ele}}_{\text{QM/MM}}$ and $E^{\text{pol}}_{\text{QM/MM}}$ are the electrostatic and polarization energy terms, respectively. Therefore, the QM-MM Pauli repulsion and dispersion interactions are usually neglected. In electrostatic QM/MM embeddings, only the electrostatic contribution in Eq. \ref{eq:en_int} is retained. In particular, each MM atom is assigned a fixed charge, which polarizes the QM density. In polarizable QM/MM approaches, \textit{both} $E^{\text{ele}}_{\text{QM/MM}}$ and $E^{\text{pol}}_{\text{QM/MM}}$ are retained. In this work, we exploit the QM/FQ approach because it is particularly tailored to treat aqueous solutions.\cite{giovannini2020csr,giovannini2020pccp} 
In QM/FQ, each atom of the MM portion is assigned a charge ($q$) which can vary in agreement with the electronegativity equalization principle (EEP),\cite{sanderson1951} i.e. a charge flow occurs when two atoms have different chemical potential.\cite{lipparini2012linear,rick1994dynamical,rick1995fluctuating,rick1996dynamical,cappelli2016integrated} The FQ force field is defined in terms of two atomic parameters, namely the electronegativity ($\chi$) and the chemical hardness ($\eta$).

In both non-polarizable QM/MM and polarizable QM/FQ, the QM/MM interaction 
energy is given by:
\begin{equation}
\label{eq:qmtip3p}
E^{\text{int}}_{\text{QM/MM}} = \sum_i q_i V_i(\mathbf{D})\ ,
\end{equation} 
where, $V_i(\mathbf{D})$ is the electric potential generated by the QM density on the i$-th$ charge ($q_i$). Notice that, in the case of QM/TIP3P, the values of the charges $q_i$ are fixed, whereas in QM/FQ they are obtained by minimizing the following energy expression:\cite{lipparini2012linear,lipparini2012analytical}
\begin{equation}
\label{eq:QMFQ}
\mathcal{E}[\mathbf{D},\mathbf{q},\boldsymbol{\mathbf{\lambda}}] =  \mbox{tr} \mathbf{h}\mathbf{D} + \frac{1}{2}\mbox{tr} \mathbf{D}\mathbf{G}(\mathbf{D}) + \frac{1}{2}\mathbf{q}_\lambda^\dagger\mathbf{M}\mathbf{q}_\lambda + \mathbf{q}_\lambda^\dagger\mathbf{C}_Q + \mathbf{q}_\lambda^\dagger\mathbf{V}(\mathbf{D}) .
\end{equation}
The term $\mathbf{q}_\lambda$ is a vector, collecting the Lagrangian multipliers ensuring charge conservation and the electric variables associated with the FQ force field, i.e. charges $q$. The $\mathbf{M}$ matrix is the interaction kernel between the charges, containing also the two Lagrangian blocks.\cite{lipparini2011polarizable,giovannini2018hyper} The term $\mathbf{q}_\lambda^{\dagger}\mathbf{C}_Q$ accounts for the interaction between permanent moments, i.e. atomic electronegativities and total charge constraints. The last term $\mathbf{q}_\lambda^\dagger\mathbf{V}(\mathbf{D})$ represents the interaction between the charges and the QM electric potential.

The FQ charges are obtained by a minimization procedure of the functional in Eq. \ref{eq:QMFQ}, which can be recasted in the following set of linear equations:
\begin{equation}
\label{eq:linearqmfq}
\mathbf{M}\mathbf{q}_{\lambda} = -\mathbf{C}_Q - \mathbf{V}(\mathbf{D}) \ .
\end{equation}
The QM Fock matrix in both QM/MM approaches is:
\begin{equation}
    F_{\mu\nu} = h_{\mu\nu} + G_{\mu\nu}(\mathbf{D}) + \sum_i q_i V_{i,\mu\nu}(\mathbf{D}) \ .
    \label{eq:fock-qmmm}
\end{equation}
Notice that in non-polarizable QM/MM, the QM/MM contribution to the Fock matrix is fixed because the charges are constant, thus it can be computed once at the beginning of the Self Consistent Field (SCF) procedure. In polarizable QM/FQ, the charges depend on the QM density, thus the QM/MM contribution to the Fock matrix has to be computed at each SCF step. In this way, the mutual polarization between the QM and FQ parts is taken into account. 


\subsection{Multilevel HF}

Multilevel Hartree-Fock (MLHF)\cite{saether2017density,hoyvik2020convergence} is a quantum-embedding approach, in which the studied system is partitioned into an active and an inactive (environment) region.
The density of the system ($\mathbf{D}$) is expressed as the sum of an active ($\mathbf{D}^A$) and an inactive ($\mathbf{D}^B$) densities:
\begin{equation}
\mathbf{D} = \mathbf{D}^A + \mathbf{D}^B.
\end{equation}
An initial set of active occupied orbitals, and consequently $\mathbf{D}^A$, is obtained through a partial limited Cholesky decomposition\cite{aquilante2006fast,sanchez2010cholesky} of $\mathbf{D}$.
An initial set of active virtual orbitals is obtained from orthonormalized projected atomic orbitals (PAOs) \cite{pulay.1983, saebo1993} centered on the atoms in the active region of the system.
The MLHF energy of the whole system can be then written as:\cite{saether2017density}
\begin{align}
E  = &\tr \textbf{hD}^A + \frac{1}{2}\tr \textbf{D}^A \textbf{G}(\textbf{D}^A) + \tr \textbf{D}^A \textbf{G}(\textbf{D}^B)  \nonumber \\
  + &\tr \textbf{hD}^B + \frac{1}{2}\tr \textbf{D}^B \textbf{G}(\textbf{D}^B) + h_{\text{nuc}}\ ,
\label{eq:mlhf_ene}
\end{align}
where $\mathbf{h}$ and $\mathbf{G}$ are the one- and two-electron integral matrices, respectively, and $h_{\text{nuc}}$ is the nuclear repulsion. In MLHF, the active density is optimized through rotations among the active orbitals, while the inactive density is fixed. The inactive density takes part in the optimization procedure entering the \textit{effective} Fock matrix expression through its two-electron integral term. The Fock matrix in the atomic orbital (AO) basis $(\{\chi_\mu\})$ reads:
\begin{equation}
\label{eq:mlhf_fock}
F_{\mu\nu} = h_{\mu\nu} + G_{\mu\nu}(\mathbf{D}^A) + G_{\mu\nu}(\mathbf{D}^B) \ .
\end{equation} 
Note that in MLHF both the electrostatic and the Pauli repulsion interactions between the active and inactive fragments are retained. 


\subsection{Multilevel Hartree-Fock AB}

The MLHF energy defined in Eq. \ref{eq:mlhf_ene} allows for the minimization of the active energy in the field generated by the inactive density. However, the different energy terms cannot be assigned to the individual fragments A and B, because they are coupled in the one-electron term $\mathbf{h}=\mathbf{T}+\mathbf{V}^A+\mathbf{V}^B$, where $\mathbf{T}$ is the kinetic operator and $\mathbf{V}^A$ and $\mathbf{V}^B$ are the electron-nuclear attraction operators for the two parts. Using the definition of $\mathbf{h}$, Eq. \ref{eq:mlhf_ene} can be written as:\cite{giovannini2020mlhf_ab}
\begin{align}
E & = \underbrace{\tr \mathbf{h}^A\mathbf{D}^A + \dfrac{1}{2}\tr \mathbf{D}^A\mathbf{G}(\mathbf{D}^A) + h^{A}_{nuc}}_{E_A} \nonumber\\
 & +\underbrace{\tr \mathbf{h}^B\mathbf{D}^B + \dfrac{1}{2}\tr \mathbf{D}^B\mathbf{G}(\mathbf{D}^B) + h^{B}_{nuc}}_{E_B} \nonumber \\
  & + \underbrace{\tr \mathbf{V}^B\mathbf{D}^A + \tr \mathbf{V}^A\mathbf{D}^B + \tr \mathbf{D}^A \mathbf{G} (\mathbf{D}^B) + h^{AB}_{nuc}}_{E_{AB}} \ ,  
 \label{eq:mlhf_energies_ab}
\end{align}
where $h^A_{nuc}$, $h^{B}_{nuc}$ and $h^{AB}_{nuc}$ are nuclear repulsion terms, and $\mathbf{h}^X = \mathbf{T} + \mathbf{V}^{X}$, with $\{X = A,B\}$. Using the Eq. \ref{eq:mlhf_energies_ab}, the active and inactive MOs can be localized in their user-specified regions. This can be done by exploiting the energy-based localization procedure recently developed by us.\cite{giovannini2020mlhf_ab}. In this approach (called MLHF-AB), the sum of $E_A$ and $E_B$ is minimized in the space spanned by the occupied orbitals of both active and inactive parts, thus the total energy defined in Eq. \ref{eq:mlhf_ene} remains constant during the procedure. The sum of the active and inactive energies reads: 
\begin{align}
E_A + E_B & = \tr \bh{A}{A} + \dfrac{1}{2}\tr \bG{A}{A} + \tr \bh{B}{} - \tr \bh{B}{A} + \nonumber \\
 & +\dfrac{1}{2}\tr \bG{}{} + \dfrac{1}{2}\tr \bG{A}{A} - \tr \bG{A}{} = \nonumber \\
 & = \tr (\mathbf{V}^A - \mathbf{V}^B) \mathbf{D}^A + \tr \bG{A}{A} - \tr \bG{A}{} \nonumber \\
 & + \tr \bh{B}{} + \dfrac{1}{2} \tr \bG{}{} \ ,
\label{eq:MLHF-Ab}
\end{align} 
where the last two terms depend on the total density $\mathbf{D}$, and are therefore constant energy terms. The Fock matrix can be written as:
\begin{align}
F_{\mu\nu} & = V^A_{\mu\nu} - V^B_{\mu\nu} + G_{\mu\nu}(\mathbf{D}^A) - G_{\mu\nu}(\mathbf{D}^B) \nonumber \\
& =V^A_{\mu\nu} - V^B_{\mu\nu} + 2G_{\mu\nu}(\mathbf{D}^A) - G_{\mu\nu}(\mathbf{D}) \ .
\end{align}
Note that minimizing the sum of A and B parts in MLHF-AB (see Eq. \ref{eq:MLHF-Ab}) is equivalent to maximizing the interaction energy $E_{AB}$, i.e. maximizing the repulsion between the two fragments. 


\subsection{CC2, MLCC2, and CC2-in-HF}
The coupled cluster wave function is given by
\begin{equation}
   \ket{\CC} = e^T \ket{\R},\quad T = \sum_{\mu}t_\mu\tau_\mu,
   \label{eq:cc-wf}
\end{equation}
where $T$ is the cluster operator, defined in terms of the cluster amplitudes, $t_\mu$, and excitation operators, $\tau_\mu$, and $\ket{R}$ is a reference determinant. The amplitudes are obtained by solving the projected coupled cluster equations,
\begin{align}
    \Omega_\mu = \bra{\mu} e^{-T} H e^{T} \ket{\R} = 0,
\end{align}
where $\ket{\mu} = \tau_\mu\ket{R}$. 
The coupled cluster ground state energy is given by:
\begin{align}
    E_\CC = \bra{\R} e^{-T} H e^{T} \ket{\R},
\end{align}
and excitation energies are calculated by determining the eigenvalues of the Jacobian matrix, 
\begin{align}
    A_{\mu\nu} = \frac{\partial\Omega_\mu}{\partial t_{\nu}}.
\end{align}
%
The CC2\cite{christiansen1995second} model is obtained by restricting $T$ to only include single and double excitations with respect to the reference,
\begin{equation}
    T = T_1 + T_2 \ ,
    \label{eq:cc2-X}
\end{equation}
and treating the double excitation operator $T_2$ perturbatively.
The projected coupled cluster equations become
\begin{align}
    \Omega_{\mu_1} & = \bra{\mu_1} \tilde{H} + [\tilde{H}, T_2] \ket{\R}\\
    \Omega_{\mu_2} & = \bra{\mu_2} \tilde{H} + [F, T_2] \ket{\R}\ ,
    \label{eq:mlcc2-projected}
\end{align}
where $\tilde{H}$ is the $T_1$-transformed Hamiltonian:
\begin{equation}
    \tilde{H} = e^{-T_1} H e^{T_1}
    \label{eq:t1-H} \ .
\end{equation}

The multilevel CC2 (MLCC2)\cite{folkestad2019multilevel} method belongs to the class of multilevel coupled cluster (MLCC) approaches, introduced by Myhre \etal \cite{myhre2013extended, myhre2014multi} 
These approaches aim to retain the accurate description of intensive properties that can be obtained with coupled cluster theory, while reducing the computational cost through the restriction of the higher order excitation operators to a subset of the molecular orbitals. 
In MLCC2, the $T_2$ operator is restricted to excite within an active orbital space, whereas the $T_1$ operator is unrestricted. If one also restricts $T_1$ to the active orbital space, one obtains the CC2-in-HF model.
The determination of a sensible active orbital space is a prerequisite for the MLCC2 and CC2-in-HF methods. Active occupied orbitals can, for instance, be constructed through a partial limited Cholesky decomposition of the Hartree-Fock density, and active virtual orbitals can be obtained as orthonormalized PAOs, as in MLHF. Alternatively, for MLCC2, correlated natural transition orbitals (CNTOs) can be employed.\cite{hoyvik2017correlated, folkestad2019multilevel} With CNTOs, no active region of the system must be selected \textit{a priori}, since the partition relies on the excitation vectors of a lower level method - i.e. CCS in a MLCC2 calculation.


\subsection{Coupling MLHF and MLCC2 with an MM layer}

In this section, we present and discuss the coupling between MLHF and MLCC2 with a third layer treated at the MM level. 
For non-polarizable MLHF/MM, this can be done by rewriting the total energy in Eq. \ref{eq:mlhf_ene} including QM/MM interaction energy terms:

\begin{align}
\label{eq:mlhf_tip3p}
E & = \tr \textbf{hD}^A + \frac{1}{2}\tr \textbf{D}^A \textbf{G}(\textbf{D}^A) + \tr \textbf{D}^A \textbf{G}(\textbf{D}^B) \nonumber \\ 
& + \tr \textbf{hD}^B + \frac{1}{2}\tr \textbf{D}^B \textbf{G}(\textbf{D}^B) + h_{nuc} + \mathbf{q}\mathbf{V}(\mathbf{D}) \ ,
\end{align}

where $\mathbf{V}(\mathbf{D})$ is the electric potential acting on each MM charge due to the total density matrix $\mathbf{D}$. The non-polarizable MLHF/MM Fock matrix is thus modified as:

\begin{equation}
\label{eq:mlhf_tip3p_fock}
F_{\mu\nu} = h_{\mu\nu} + G_{\mu\nu}(\mathbf{D}^A) + G_{\mu\nu}(\mathbf{D}^B) + \mathbf{q}\mathbf{V}_{\mu\nu} \ ,
\end{equation} 
which corresponds to Eq. \ref{eq:mlhf_fock}, with the additional $\mathbf{q}\mathbf{V}_{\mu\nu}$ term that is independent of the active density matrix. Therefore, such a contribution can be calculated once at the beginning of the SCF procedure. 

In case of QM/FQ the total MLHF/FQ energy instead reads:
\begin{align}
\label{eq:mlhf_FQ}
E & = \tr \textbf{hD}^A + \frac{1}{2}\tr \textbf{D}^A \textbf{G}(\textbf{D}^A) + \tr \textbf{D}^A \textbf{G}(\textbf{D}^B) \nonumber \\ 
& + \tr \textbf{hD}^B + \frac{1}{2}\tr \textbf{D}^B \textbf{G}(\textbf{D}^B) + h_{nuc}   \nonumber \\ & +\frac{1}{2}\mathbf{q}_\lambda^\dagger\mathbf{M}\mathbf{q}_\lambda + \mathbf{q}_\lambda^\dagger\mathbf{C}_Q + \mathbf{q}_\lambda^\dagger\mathbf{V}(\mathbf{D}) \ .
\end{align}
In order to calculate the FQ charges, Eq. \ref{eq:linearqmfq} can be used, by setting $\mathbf{D} = \mathbf{D}^A + \mathbf{D}^B$, where $\mathbf{D}^B$ is constant during the SCF procedure. The QM/FQ Fock matrix reduces to Eq. \ref{eq:mlhf_tip3p_fock}, however the charges vary during the SCF procedure, thus the last term is calculated at each step. In this way, mutual polarization between the MLHF active part and FQ charges is taken into account.
The coupling between MLCC2 and a third MM layer is instead trivial because the MM contributions are introduced at the SCF level only. Therefore, if the reference state is calculated at the HF/MM level, the coupling between MLCC2 and the MM layer is automatically included. 

\section{Computational Details}

The CC2-in-MLHF/MM, CC2-in-MLHF-AB/MM, CC2-in-HF/MM and MLCC2/MM methods are implemented in a development version of the electronic structure program $e^{T}$.\cite{eT2020} The MLHF calculations are performed by first diagonalizing the initial Fock matrix constructed by using a
superposition of molecular  densities.\cite{neugebauer2005merits,he2010divide,marrazzini2020mldft} Then, the obtained density is partitioned into $A$ and $B$ densities, using Cholesky decomposition (threshold: 10$^{-2}$) and PAOs for active occupied and virtual orbitals, respectively.\cite{aquilante2006fast,sanchez2010cholesky,saether2017density,pulay.1983, saebo1993} Finally, the MLHF energy in Eq. \ref{eq:mlhf_ene} is minimized in the active MO space. The MLHF-AB/MM calculations are performed using the procedure developed in Ref. \citenum{giovannini2020mlhf_ab}. In particular, the MLHF and occupied-occupied rotations minimizing Eq. \ref{eq:mlhf_energies_ab} are iterated until convergence. In the MLHF-AB calculations, the number of Cholesky diagonals corresponds to the number of occupied orbitals $n_o$ for the active atoms, i.e. the largest $n_o$ diagonals of the density matrix are selected.\cite{giovannini2020mlhf_ab}
For both MLHF and MLHF-AB, the excitation energies are computed by means of a CC2 calculation (CC2-in-MLHF(-AB)) performed in the same orbital space defined as MLHF(-AB) active.

The MLCC2 calculations are instead performed in the framework of the CNTOs approximation. These orbitals are tailored for the excited state calculation performed, specifying the corresponding state numbers. The number of occupied CNTOs is requested as well, while the number of virtual is automatically defined as $n_v = n_{o}\frac{N_v}{N_o}$, where the lowercase letters refer to the active virtual and occupied sets and the uppercase letters to full space.\cite{folkestad2019multilevel} The double excitation operator indices are then restricted to the active orbitals.

In addition to the aforementioned methods we also perform CC2-in-HF calculations, where only a subspace of the HF space is treated at CC2 level. A limited Cholesky decomposition (threshold: 10$^{-2}$) is used to define the occupied orbitals, while the virtual orbitals are separated through orthogonalized PAOs.

\begin{figure}[htbp!]
\centering
\includegraphics[width=.4\textwidth]{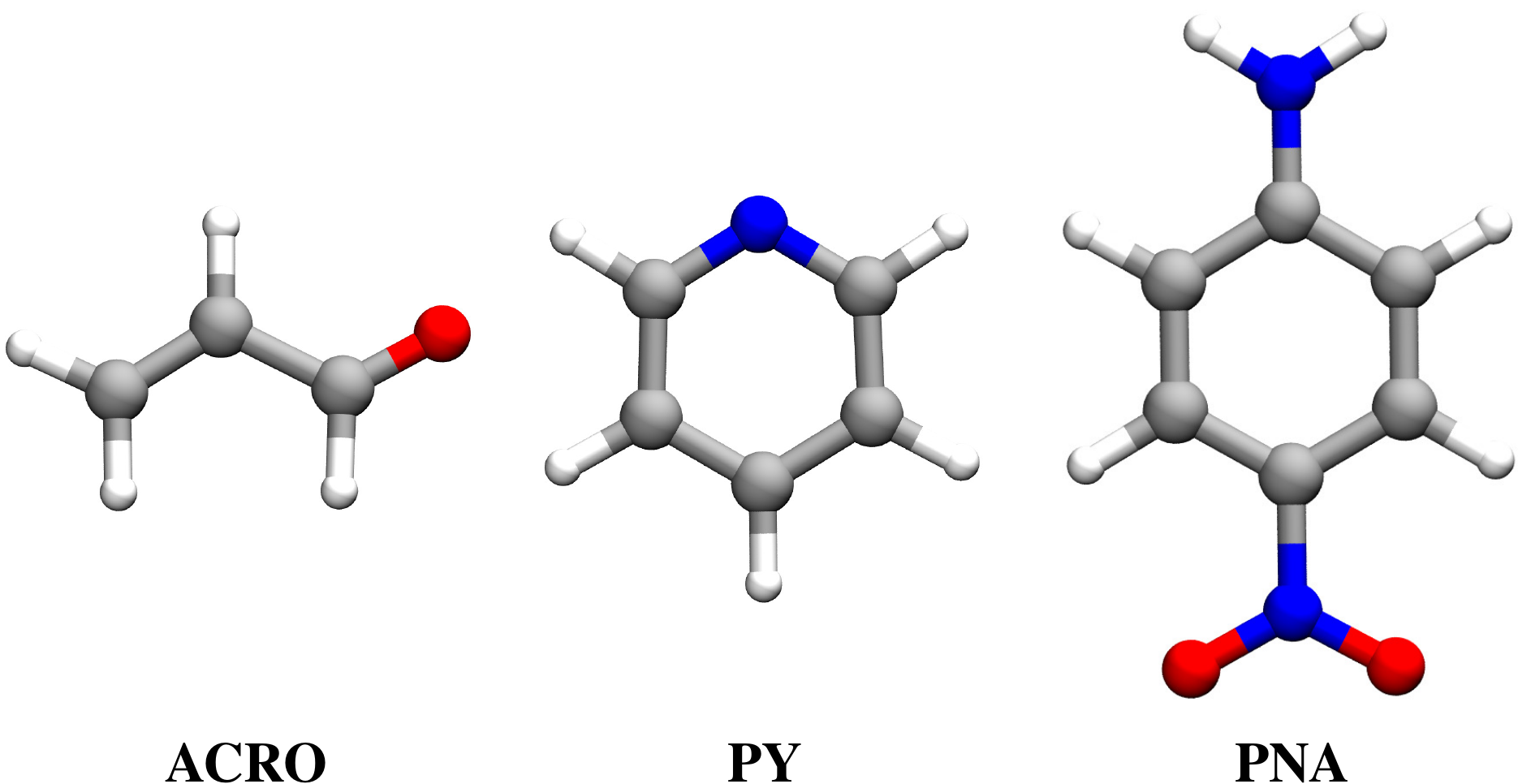}
\caption{Molecular structures of ACRO (left), PY (middle) and PNA (right).}
\label{fig:molecules_structures}
\end{figure}

All the discussed methods are applied to the calculations of excitation energies of acrolein (ACRO), pyridine (PY) and para-nitroaniline (PNA) in aqueous solution. Such molecules are chosen because their vacuo-to-water solvatochromism has been amply studied both theoretically and experimentally.\cite{sok2011solvent,kosenkov2010solvent,eriksen2013failures,sneskov2011scrutinizing,olsen2010excited,eriksen2012importance,frutos2013theoretical,defusco2011modeling,mennucci2002hydrogen,pagliai2017electronic,cossi2001time,mason1959electronic,millefiori1977electronic,cai2000low,schreiber2008benchmarks,prabhumirashi1986solvent,kovalenko2000femtosecond,de2001monte,moskvin1966experimental,aidas2008performance,aquilante2003theoretical,brancato2006quantum,marenich2011practical,duchemin2018bethe,bistafa2016complete} In order to take into account the specific and dynamical aspects of solvation, we follow the computational protocol reported in Ref. \citenum{giovannini2020csr}, which is designed to accurately describe the particular case of aqueous solutions. More specifically, for each molecule a classical molecular dynamics (MD) simulation was performed according to the procedure reported in the literature.\cite{giovannini2019fqfmulinear,giovannini2019quantum} A set of 50 uncorrelated snapshots is extracted from the MD and a solute-centered sphere is cut. The radius of the sphere is chosen to retain all solute-solvent interactions, and it is set to 15 \AA~ (containing approximately 400 water molecules).\cite{giovannini2019fqfmulinear,giovannini2019quantum}

On the obtained droplets, we calculate CC2/aug-cc-pVDZ excitation energies of the solute moieties using the frozen core approximation and the different methods exposed above. In particular, the QM regions are defined by selecting the solute and all the water molecules that are placed at a distance lower than 3.5 \AA~ with respect to each solute atom. In this way, the first solvation sphere is included.\cite{giovannini2019quantum,giovannini2019fqfmulinear} 
All the remaining water molecules are described using the non-polarizable TIP3P force field\cite{mark2001structure} (QM/TIP3P) or the polarizable FQ force field\cite{giovannini2020csr} (QM/FQ), with the parametrization proposed in Ref. \citenum{giovannini2019eprdisrep}. 

The default thresholds of the $e^{T}$ program\cite{eT2020} are used, with the exception of the residual threshold in the CC excited state equations and of the Cholesky solver decomposition threshold,\cite{folkestad2019cholesky} both set to 10$^{-4}$. The excitation energy of the state $s$ given a set of $N_{snap}$ snapshots is calculated as:\cite{marenich2014electronic}
\begin{equation}
    <\omega^{s}>_{MD} = \dfrac{\sum_i^{N_{snap}} f^s_i \omega^s_i}{\sum^{N_{snap}}_i f^s_i} \ ,
    \label{eq:w_md}
\end{equation}
where, $f^s_i$ and $\omega^s_i$ are the oscillator strength and the excitation energy of the $i$-th snapshot, respectively. 

The current implementation of MLCC2 in $e^{T}$ is limited to the calculation of excitation energies. Therefore, for all MLCC2 calculations, we use the CC2-in-HF oscillator strengths in Eq. \ref{eq:w_md}. 

\section{Numerical Results}

The selected molecules are characterized by 
$n \rightarrow \pi^*$ (PY), $\pi\rightarrow\pi^*$ (PNA) and both $n \rightarrow \pi^*$ and $\pi\rightarrow\pi^*$ (ACRO)
electronic transitions. 
In this section, we first report a benchmark of the different theoretical methods on a single randomly selected snapshot of ACRO, as extracted from the MD simulation. Then, the methods that are found to provide a good compromise between the accuracy and the computational cost are applied to the full set of extracted snapshots of ACRO, PY and PNA. The chosen approaches are compared to the experimental results taken from the literature.\cite{marenich2014electronic,aidas2008performance}

\subsection{Benchmark}

The performances of all the different combinations of the developed theoretical approaches are tested for a randomly selected snapshot of ACRO in aqueous solution. Acrolein is selected because its small size allows for a comparison with high level calculations.

\begin{figure}[htbp!]
\centering
\includegraphics[width=.5\textwidth]{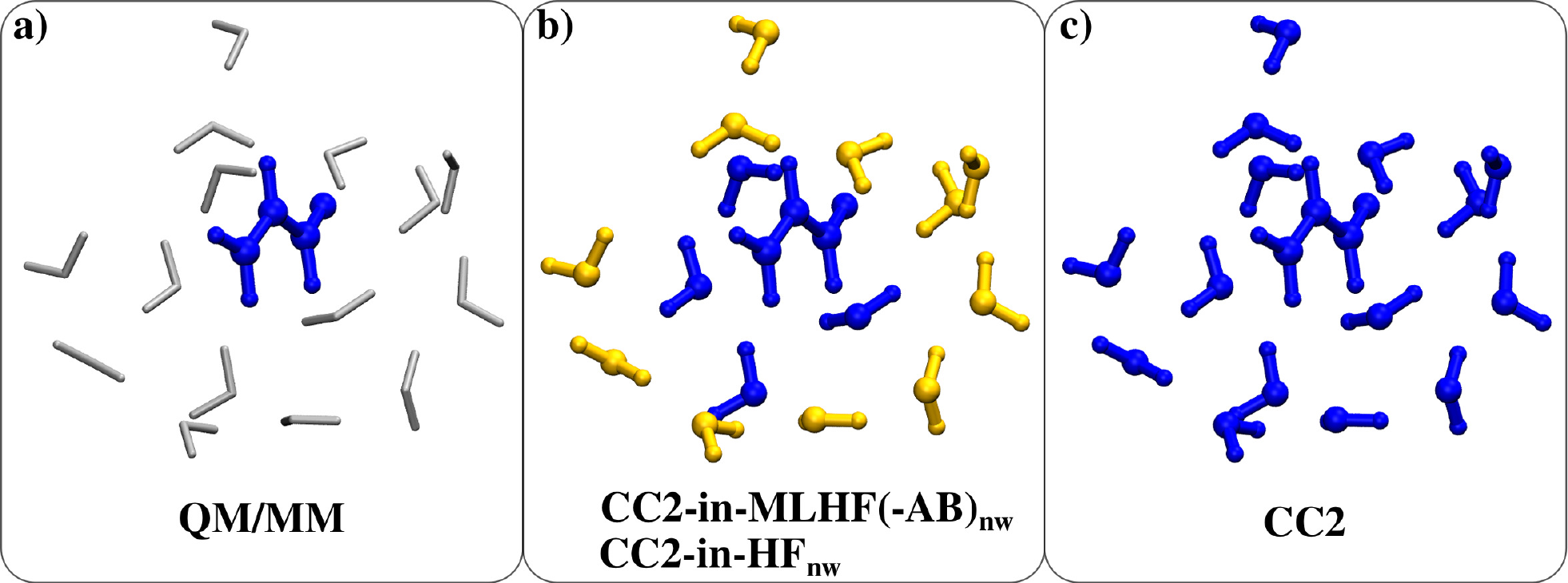}
\caption{Graphical depiction of the reduced snapshot of ACRO in aqueous solution as partitioned by (\textbf{a}) QM/MM approaches; (\textbf{b}) CC2-in-MLHF$_{\text{nw}}$, CC2-in-MLHF-AB$_{\text{nw}}$ and CC2-in-HF$_{\text{nw}}$ (the active part is represented in blue, the inactive one in yellow); (\textbf{c}) 
CC2.}
\label{fig:acro_12_small_structure}
\end{figure}

First, we study the performances of the different methods on a reduced snapshot obtained by retaining all the water molecules that are closer than 3.5 \AA~ from each acrolein atom, and by removing the remaining water molecules. This results in a snapshot with ACRO and 15 water molecules. The obtained system is described at different levels of theory:
\begin{itemize}
\item QM/MM: the ACRO moiety is treated at the CC2 level, whereas the 15 water molecules are described either at the TIP3P (i.e. electrostatic embedding) or at the FQ level (see Fig. \ref{fig:acro_12_small_structure} a). 
\item 
CC2-in-MLHF(-AB): only the ACRO moiety is treated as the active fragment at the MLHF level, whereas the 15 water molecules are inactive (see Fig. \ref{fig:acro_12_small_structure} b).  
\item 
CC2-in-MLHF(-AB)$_{\text{nw}}$: 
the ACRO moiety and the water molecules closest to the ACRO center of mass are included in the active MLHF part. We include up to 5 water molecules (see Fig. \ref{fig:acro_12_small_structure} b). The $n$ in the acronym specifies the number of water molecules included in the active part, while the remaining water molecules are inactive. 
\item 
MLCC2$_{\text{nCNTOs}}$: we select a number $n$ of active occupied CNTOs equal to 25, 30, 35. In this way, we select the number of occupied orbitals present in ACRO and two, three or four water molecules. 
\item CC2-in-HF$_{\text{nw}}$: the reduced snapshot is treated with HF, and only ACRO and $\text{n}$ water molecules are active in the CC2 calculation (see Fig. \ref{fig:acro_12_small_structure} b).
\item CC2: the reduced snapshot is fully described at the CC2/aug-cc-pVDZ level of theory. This is the reference value for all the used methods (see Fig. \ref{fig:acro_12_small_structure} c).
\end{itemize}

\begin{figure}[htbp!]
\centering
\includegraphics[width=.4\textwidth]{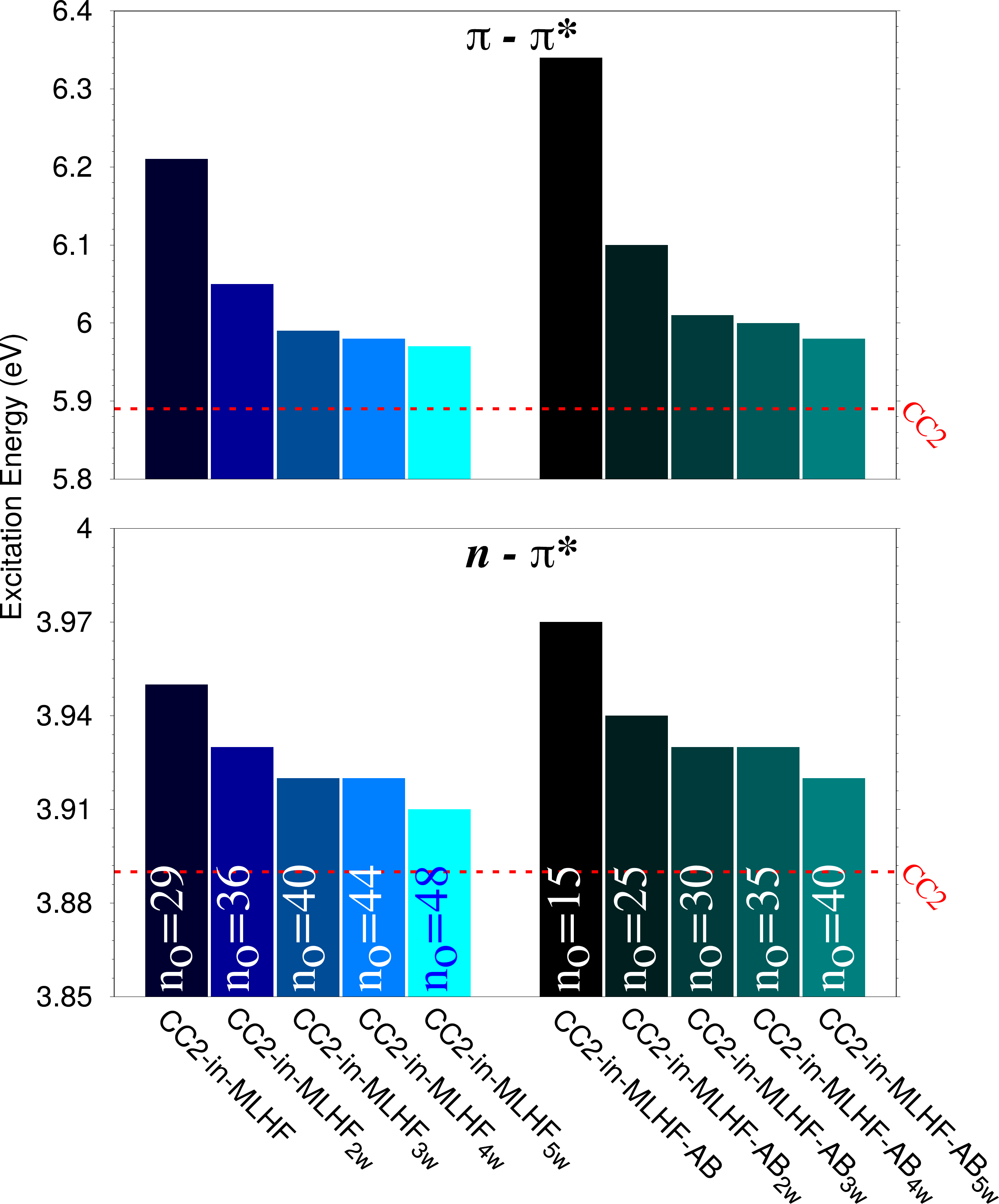}
\caption{The $n - \pi^*$ and $\pi - \pi^*$ excitation energies of the reduced snapshot of ACRO in aqueous solution computed at the CC2/aug-cc-pVDZ level, using the MLHF (left) and MLHF-AB (right) reference states.}
\label{fig:acro_12_small_results_mlhf}
\end{figure}

\begin{figure}[htbp!]
\centering
\includegraphics[width=.4\textwidth]{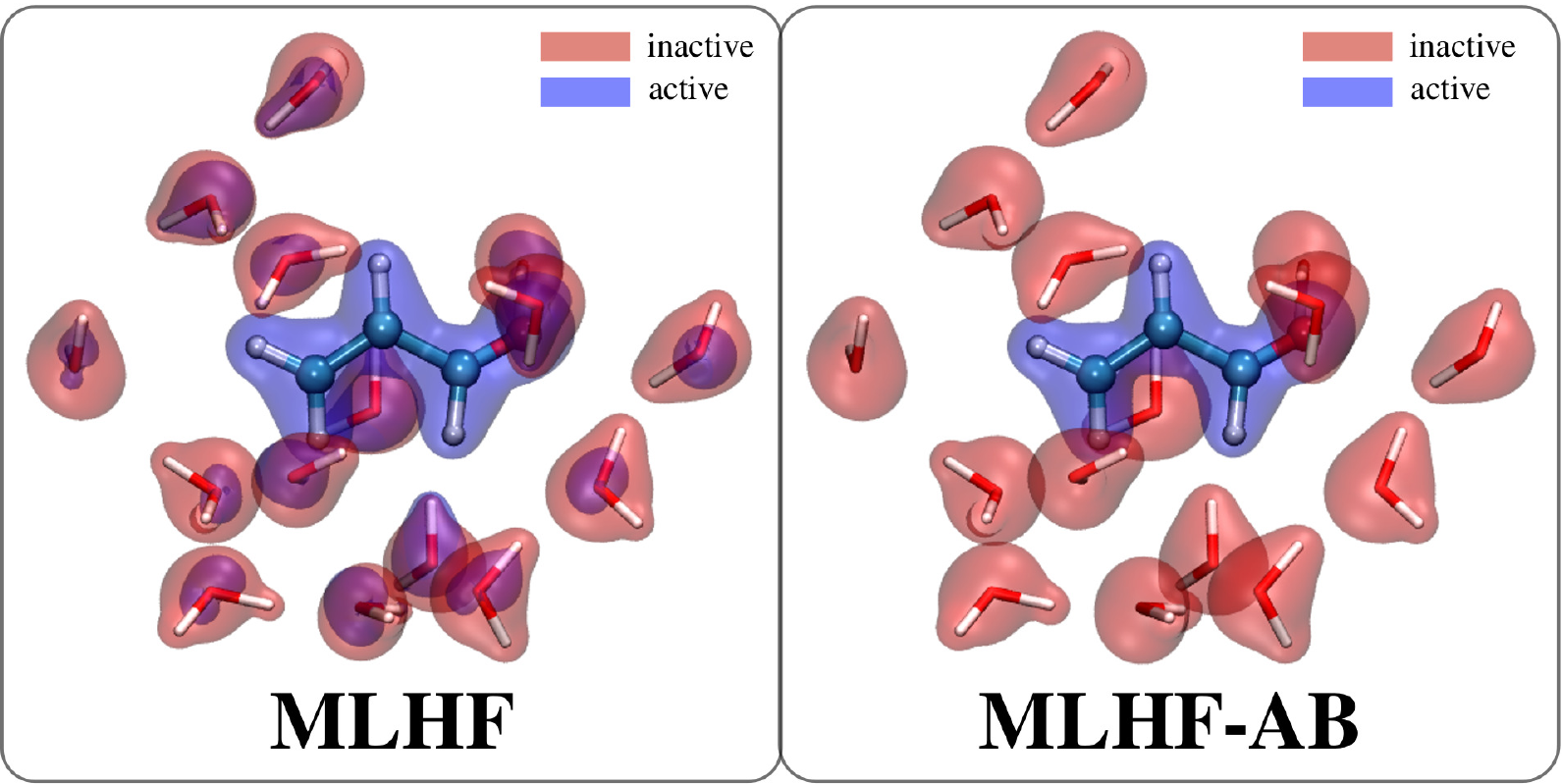}
\caption{MLHF and MLHF-AB active (blue) and inactive (red) HF densities of the reduced snapshot with ACRO being the required active part.}
\label{fig:density_acro_mlhf}
\end{figure}

We first comment on the results obtained using MLHF or MLHF-AB to describe the ground state reference. The main difference between the two approaches is that the number of occupied active orbitals in MLHF is determined by the Cholesky threshold, which is set to 10$^{-2}$, according to previous studies.\cite{myhre2014multi,folkestad2020multilevel} In MLHF-AB, 
the number of occupied orbitals centered on the active atoms 
is instead selected. MLHF and MLHF-AB are applied to the small snapshot represented in Fig. \ref{fig:acro_12_small_structure}b, treating ACRO and $n$ water molecules as active, whereas the remaining water molecules are inactive. The CC2-in-MLHF and CC2-in-MLHF-AB $n-\pi^*$ and $\pi-\pi^*$ excitation energies are reported in Fig. \ref{fig:acro_12_small_results_mlhf}. The two approaches mostly differ for the results reported when only the solute is treated as active, with CC2-in-MLHF-AB predicting an higher excitation energy for both $n-\pi^*$ (0.02 eV) and $\pi-\pi^*$ (0.13 eV) transitions. This is due to the fact that the number of occupied orbitals selected in MLHF 
is almost twice the number of those used in MLHF-AB (29 vs. 15). To better understand the differences between the CC2-in-MLHF and CC2-in-MLHF-AB methods, in Fig. \ref{fig:density_acro_mlhf} we report the computed ground state active and inactive densities. As it can be seen, with the MLHF reference the active density is spread also into the inactive region, whereas with MLHF-AB it is well localized in the ACRO moiety only. This is the origin of the observed differences in the excitation energies reported in Fig. \ref{fig:acro_12_small_results_mlhf}.

The discrepancy between the two methods rapidly decreases as a larger number of water molecules is included in the active region. In particular, CC2-in-MLHF-AB$_{5w}$ excitation energies only 
differ by 0.01 eV with respect to the results obtained by means of the CC2-in-MLHF$_{5w}$ approach (see also Table S1 given in the Electronic Supplementary Information - ESI). However, if the same active atoms are requested, the number of occupied in MLHF-AB are almost 10 less than MLHF. This yields a reduction of the computational cost which ranges from 50\% to  33\% going from zero to five water molecules included in the active part. 

\begin{figure}[htbp!]
\centering
\includegraphics[width=.5\textwidth]{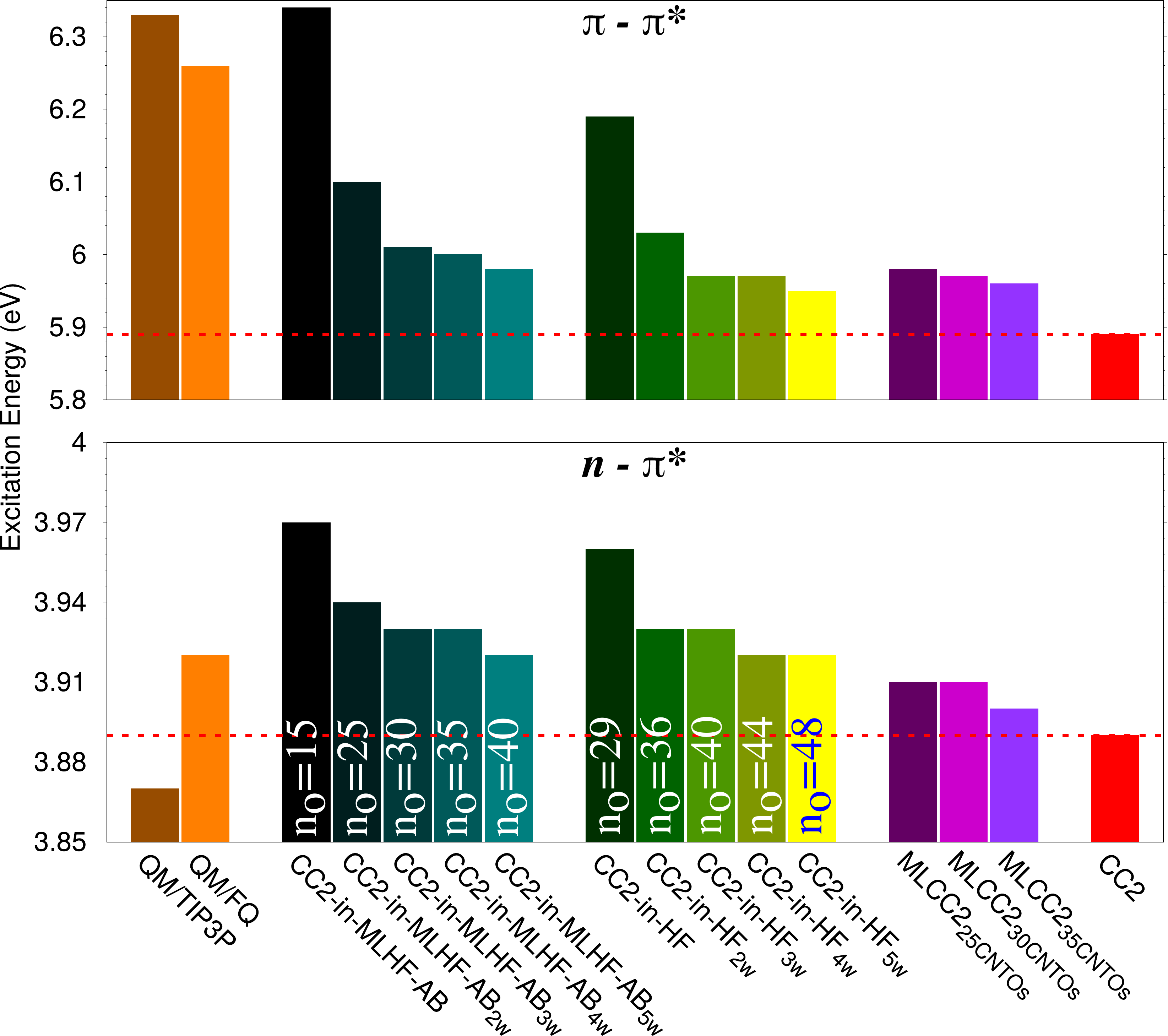}
\caption{The $n - \pi^*$ and $\pi - \pi^*$ excitation energies of the reduced snapshot of ACRO in aqueous solution (see Fig. \ref{fig:acro_12_small_structure}) computed at CC2/aug-cc-pVDZ level of theory, using different approaches to restrict the orbital space.}
\label{fig:acro_12_small_results}
\end{figure}

We now move on to comment on the results obtained by using all the other approaches, which are reported in Fig. \ref{fig:acro_12_small_results} and in Tab. S1 in the ESI. We recall that the full space CC2
result is used as reference (see the red line in Fig. \ref{fig:acro_12_small_results}). Both QM/TIP3P and QM/FQ approaches are not able to correctly reproduce the reference value for either $n-\pi^*$ or $\pi-\pi^*$ excitations. These results are in line with what has already been reported in the literature. In fact, Pauli repulsion effects, which are neglected in the considered methods, play a crucial role in the correct description of ACRO transitions.\cite{giovannini2019quantum,aidas2008performance} Such a confinement effect is taken into account by all the quantum-embedding approaches used in this work, i.e. CC2-in-MLHF(-AB), CC2-in-HF and MLCC2. In particular, by enlarging the active part in the CC2-in-MLHF(-AB) 
and CC2-in-HF methods, the excitation energies of both studied transitions approach the reference CC2 values. The best agreement is clearly achieved by including 5 water molecules in the active region; however, a good compromise between computational cost and accuracy is also reached by including 3 to 4 water molecules. The MLCC2 approach, finally, gives the best agreement with the reference data, and the results are almost constant by considering 25, 30 or 35 active CNTOs. However, we note that an increase in the number of active CNTOs will converge to the reference value. Moreover, whereas in all the other exploited multilevel methods the $n$ active water molecules are selected on a geometrical criterion, in MLCC2 the active-inactive partitioning is performed based on the CCS excitation vectors.

\begin{figure}[htbp!]
\centering
\includegraphics[width=.5\textwidth]{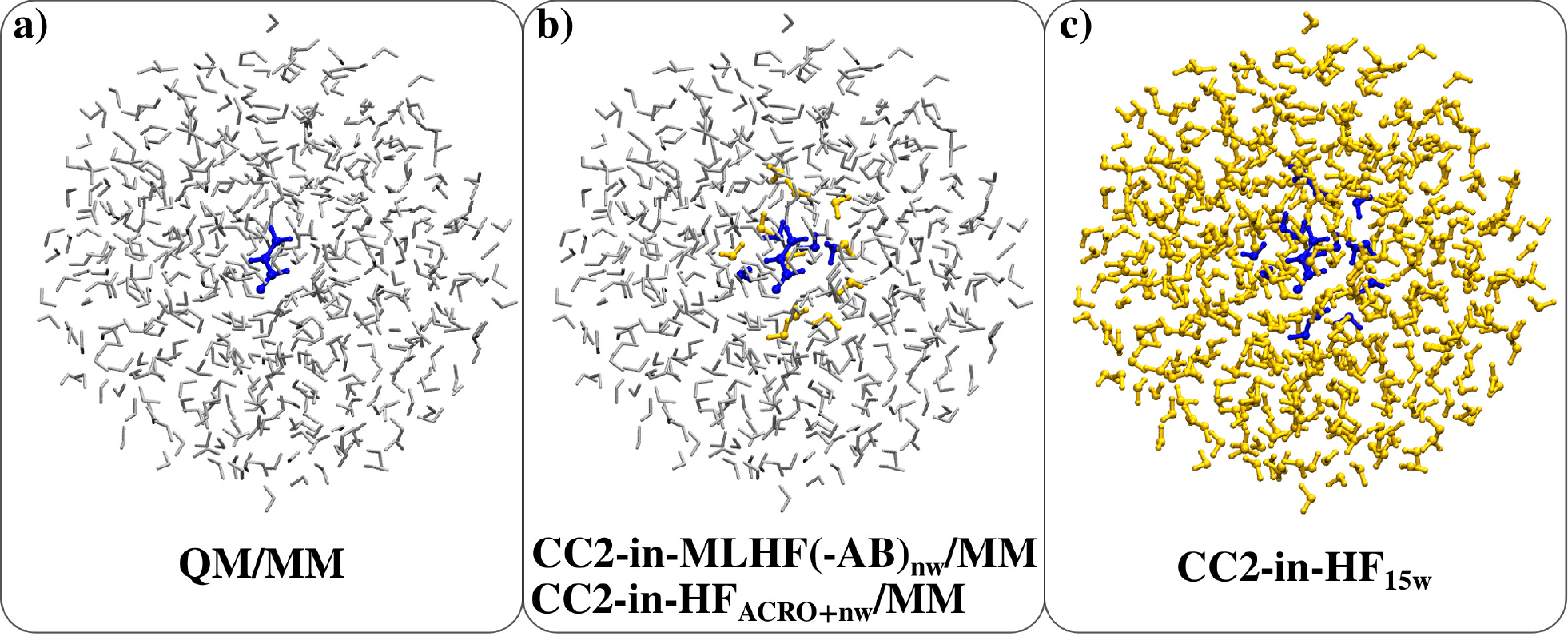}
\caption{Graphical depiction of the full snapshot of ACRO in aqueous solution as partitioned by (\textbf{a}) 
QM/MM approaches; (\textbf{b}) 
CC2-in-MLHF$_{\text{nw}}$/MM, CC2-in-MLHF-AB$_{\text{nw}}$/MM and CC2-in-HF$_{\text{nw}}$/MM, (the active part is represented in blue, the inactive one in yellow, the MM portion in silver licorice); (\textbf{c}) CC2-in-HF$_{\text{15w}}$ (CC2 active in blue, HF active: blue + yellow).}
\label{fig:acro_12_big_structure}
\end{figure}

All the investigated methods are also applied to the full snapshot, i.e. the previously discarded water molecules are included in the calculations, and treated in all cases by means of the non-polarizable TIP3P\cite{mark2001structure} or the polarizable FQ force field. This allows for the study of the performances of the three-layer approaches (CC2-in-MLHF(-AB)/MM, CC2-in-HF/MM, MLCC2/MM) we have developed. In particular, the long-range interactions are retained at the electrostatic level (with polarization effects in case of QM/FQ), 
thus neglecting Pauli repulsion. This is justified by the fact that Pauli repulsion is a short-range interaction. For the full snapshot, the reference approach is represented by the CC2-in-HF$_{\text{15w}}$
results. In the latter method, the whole snapshot is treated at the HF level, whereas the small subsystem constituted by ACRO and 15 water molecules is CC2 active. A schematic picture of the partitioning used in the different methods is shown in Fig. \ref{fig:acro_12_big_structure}.

\begin{figure}[htbp!]
\centering
\includegraphics[width=.5\textwidth]{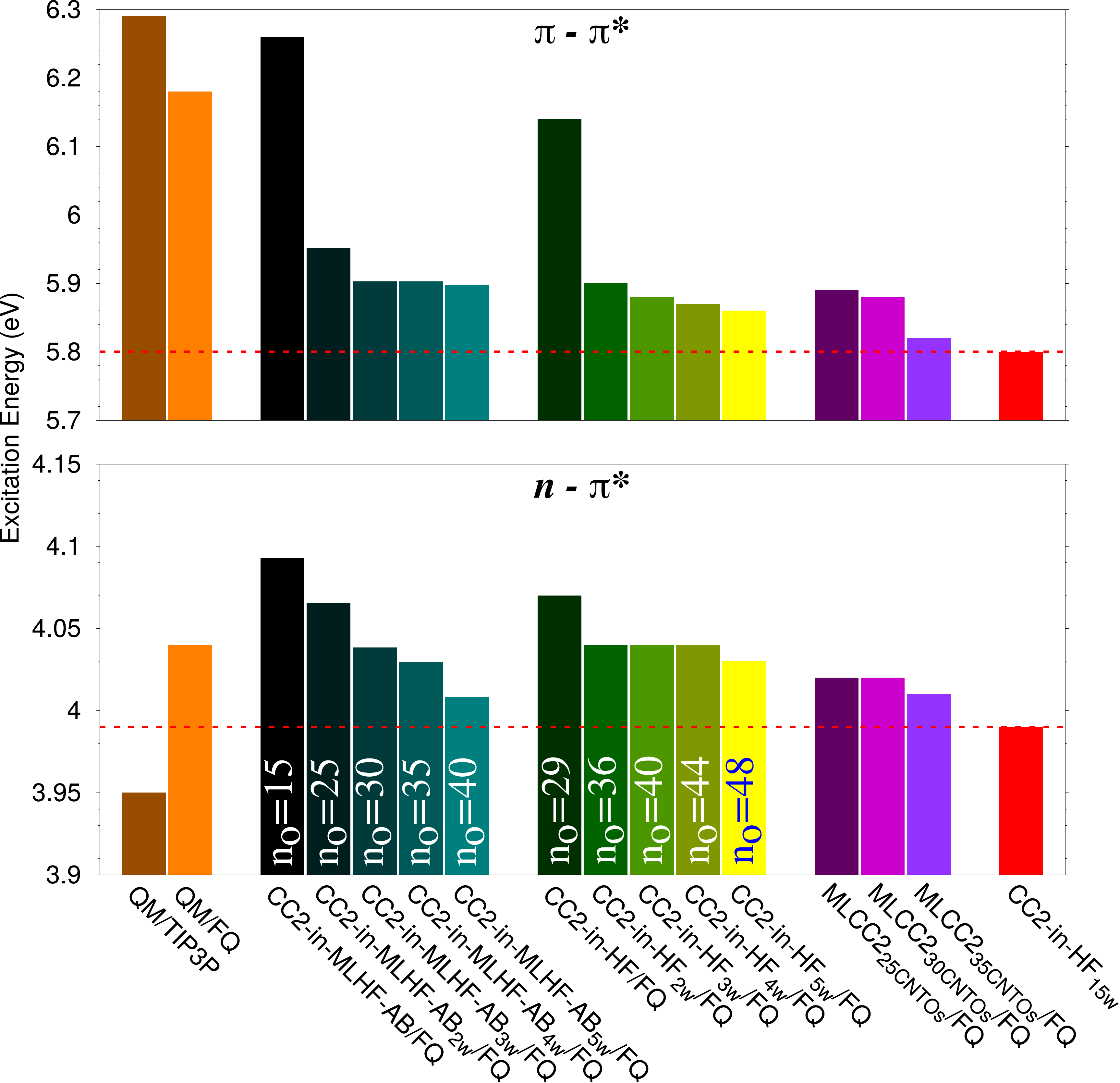}
\caption{The $n - \pi^*$ and $\pi - \pi^*$ excitation energies of the full snapshot of ACRO in aqueous solution (see Fig. \ref{fig:acro_12_big_structure}) computed at CC2/aug-cc-pVDZ level of theory, using different approaches to describe the reference state.}
\label{fig:acro_12_big_results}
\end{figure}

The $n-\pi^*$ and $\pi-\pi^*$ excitation energies obtained using 
QM/TIP3P, QM/FQ, CC2-in-MLHF-AB/FQ, CC2-in-HF/FQ, MLCC2 and CC2-in-HF$_\text{15w}$ are reported in Fig. \ref{fig:acro_12_big_results} and in Tab. S2 in the ESI. Note that we also perform a comparison between CC2-in-MLHF-AB/FQ and CC2-in-MLHF/FQ, as previously presented in Fig. \ref{fig:acro_12_small_results_mlhf} (see Fig. S1 given in ESI). From the comparison between the reference values reported in Fig. \ref{fig:acro_12_big_results} and \ref{fig:acro_12_small_results}, we notice that the inclusion of the previously discarded water molecules leads to an opposite shift for the two transitions. In particular, the CC2-in-HF$_\text{15w}$ $n-\pi^*$ excitation energy increases of almost 0.1 eV with respect to the CC2 
value of the reduced structure; for the bright $\pi-\pi^*$ excitation, the absorption energy is redshifted of about 0.1 eV. 

The QM/TIP3P approach is not able to reproduce such shifts with respect to the reduced structure. In fact, the $n-\pi^*$ excitation energy increases of almost 0.08 eV, whereas the $\pi-\pi^*$ value remains almost unchanged (-0.04 eV). On the other hand, both excitations are correctly shifted when the polarizable QM/FQ model is used (see Tabs. S2-S3 in the ESI). Therefore, the polarization effects need to be taken into account to correctly reproduce electrostatic long-range effects. For this reason, we will present the results obtained by using the different QM approaches (CC2-in-MLHF, CC2-in-MLHF-AB, CC2-in-HF, 
MLCC2) coupled with the polarizable FQ approach. The results obtained by coupling the quantum-embedding approaches with the non-polarizable TIP3P are given in Tab. S3 in the ESI.

The results reported for all the methods coupled with the FQ force field confirm the correct description provided by FQ in the prediction of the shifts with respect to the reduced snapshots (see Tab. S2 in the ESI). Moreover, the general trends already discussed with reference to Fig. \ref{fig:acro_12_small_results} continue to be valid also in this case. A good agreement with the CC2-in-HF$_{15w}$ reference is obtained when 3 to 4 water molecules are included in the active part of CC2-in-MLHF-AB/FQ and CC2-in-HF/FQ approaches. Finally, the MLCC2 provides the best agreement with the reference data, and the results are almost constant when considering 25, 30 or 35 active occupied CNTOs, with the lowest discrepancy shown by the MLCC2$_{\text{35CNTOs}}$/FQ.

Based on this benchmark, we have selected six different approaches to be used in the calculation of the excitation energies of ACRO, PY and PNA in aqueous solution. In particular, we will present results obtained by exploiting the non-polarizable QM/TIP3P and different combinations of polarizable QM/FQ methods. In both QM/TIP3P and QM/FQ, only the solvated solutes are treated at the CC2 level. CC2-in-MLHF(-AB)/FQ calculations are performed by including three water molecules in the active part, whereas in CC2-in-HF, four 
active water molecules are selected. For MLCC2/FQ calculations, a number of active CNTOs corresponding to the number of occupied 
orbitals of the solute and three water molecules is selected. The final results are obtained as an average over a set of representative structures extracted from classical MD simulations.

\subsection{ACRO in aqueous solution}

\begin{table*}[htbp!]
\centering
\begin{tabular}{lcc|cc}
\hline
       & \multicolumn{2}{c}{$n - \pi^*$} & \multicolumn{2}{c}{$\pi - \pi^*$} \\
Method & Exc. Ene. (eV) & Shift (eV) & Exc. Ene. (eV) & Shift (eV) \\
\hline
vacuo                           & 3.86   &    --   & 6.78 &  --\\
\hline
exp (vac)                      &  3.69 & -- & 6.42 & --  \\
\hline
\hline
QM/TIP3P                        &  4.11 $\pm$ 0.02 & -0.25 $\pm$ 0.02 & 6.39 $\pm$ 0.03 & 0.39 $\pm$ 0.03 \\
QM/FQ                           &  4.26 $\pm$ 0.02 & -0.40 $\pm$ 0.02 & 6.23 $\pm$ 0.03 & 0.55 $\pm$ 0.03 \\
CC2-in-MLHF$_{\text{3w}}$/FQ      &  4.13 $\pm$ 0.02 & -0.27 $\pm$ 0.02 & 6.19 $\pm$ 0.03 & 0.59 $\pm$ 0.03 \\
CC2-in-MLHF-AB$_{\text{3w}}$/FQ   &  4.16 $\pm$ 0.02 & -0.30 $\pm$ 0.02 & 6.26 $\pm$ 0.03 & 0.52 $\pm$ 0.03 \\
CC2-in-HF$_{\text{4w}}$/FQ &  4.17 $\pm$ 0.02 & -0.31 $\pm$ 0.02 & 6.21 $\pm$ 0.03 & 0.57 $\pm$ 0.03 \\
MLCC2$_{\text{30CNTOs}}$/FQ     &  4.12 $\pm$ 0.02 & -0.26 $\pm$ 0.02 & 6.14 $\pm$ 0.03 & 0.64 $\pm$ 0.03 \\
\hline
exp (wat)                      &  3.94 & -0.25 & 5.90 & 0.52  \\
\hline
\end{tabular}
\caption{$n-\pi^*$ and $\pi-\pi^*$ excitation energies and vacuo-to-solvent solvachromic shifts of ACRO in vacuo and in aqueous solution. Theoretical solvatochromic shifts are calculated with respect to the vacuo value. Computed standard errors at the 68\% confidence interval are also given. Experimental data are reproduced from Ref. \citenum{aidas2008performance}.}
\label{tab:acro_w}
\end{table*}

\begin{figure}[htbp!]
\centering
\includegraphics[width=.4\textwidth]{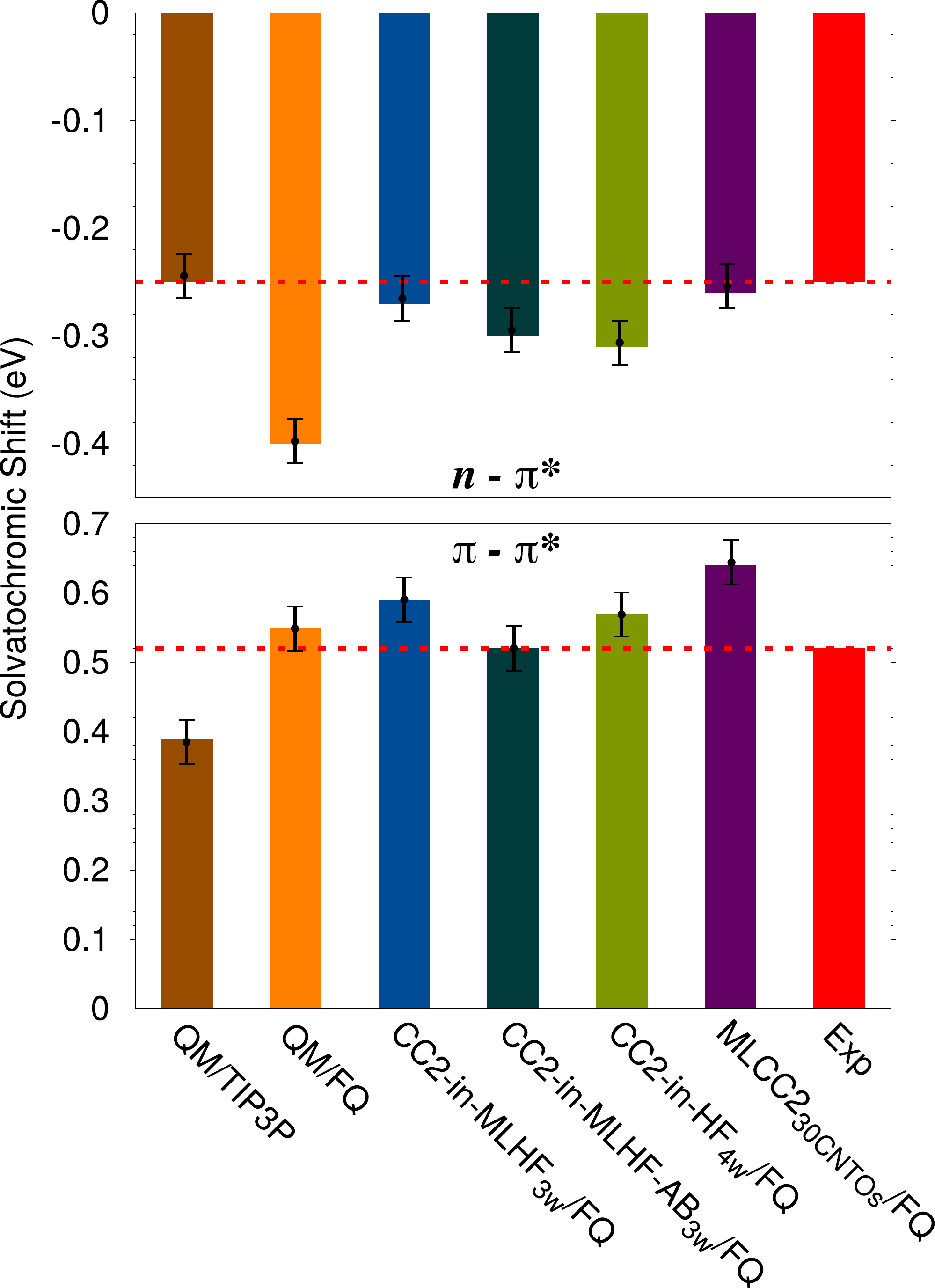}
\caption{Computed and experimental $n - \pi^*$ (top) and $\pi - \pi^*$ (bottom) vacuo-to-water solvatochromic shifts of ACRO (values 
are given in Tab. \ref{tab:acro_w}). Computed standard error bars 
at the 68\% confidence interval are also plotted. The horizontal dashed lines follow the experimental data reproduced from Ref. \citenum{aidas2008performance}.}
\label{fig:acro_MD}
\end{figure}

The first system we have studied is acrolein (ACRO) in aqueous solution. As discussed above, such a system is characterized by both a dark $n - \pi^*$ and a bright $\pi - \pi^*$ transitions. They present opposite solvatochromic shifts when passing from the gas-phase to the water solution.\cite{aidas2008performance} In particular, the vacuo-to-solvent experimental $n - \pi^*$ solvatochromic shift 
consists in about -0.25 eV, i.e. the excitation energy increases. On the other hand, the $\pi - \pi^*$ absorption energy decreases
when ACRO is dissolved in water, yielding a positive solvatochromism of 0.52 eV.\cite{aidas2008performance} Although ACRO is a relatively small system, the theoretical prediction of the experimental data is particularly challenging, because quantum effects, such as Pauli repulsion, must be taken into account.\cite{giovannini2019quantum,aidas2008performance} With the exception of QM/MM methods, all the other quantum-embedding approaches we have benchmarked in the previous section can effectively reproduce the quantum confinement effects due to Pauli repulsion. The calculated excitation energies and the vacuo-to-solvent solvatochromic shifts are reported in Tab. \ref{tab:acro_w}, together with their experimental counterparts. The solvatochromic shifts are also graphically illustrated in Fig. \ref{fig:acro_MD}. 
The gas-phase excitation energies have been calculated with a geometry optimized at the B2PLYP/cc-pVTZ level. We first comment on the trends of the excitation energies when moving from the gas phase to the aqueous solution. For the $n-\pi^*$ transition, all the investigated methods show a shift to higher energies with respect to 
the vacuo value; the opposite is the case for the $\pi-\pi^*$ excitation. 

Comparing the gas-phase computed values with their experimental counterparts, a discrepancy of almost 0.3 eV is observed. This is due to the level of theory (CC2/aug-cc-pVDZ) which was used in all the calculations. Although such a difference is quite large, we point out that we are interested in the calculation of the vacuo-to-water solvatochromic shift, which, being a difference, is less affected by absolute errors of the excitation energies.\cite{loco2016qm} The solvatochromic shifts are reported both in Tab. \ref{tab:acro_w} and in Fig. \ref{fig:acro_MD}. In agreement with the experimental data, the computed values show an opposite vacuo-to-water solvatochromic shift, being positive for $\pi-\pi^*$, and negative for $n-\pi^*$. 

The results obtained for QM/TIP3P and QM/FQ show a good agreement only with one of the two studied transitions --- $n-\pi^*$ and $\pi-\pi^*$, respectively. This behavior has already been reported at the TD-DFT level in Ref. \citenum{giovannini2019quantum}, in which it is also shown that the inclusion of Pauli repulsion, which is neglected in common QM/MM approaches, deteriorates the agreement with the experiment. Therefore, the almost perfect agreement reported for QM/MM methods is primarily due to error cancellation. 

Pauli repulsion contributions, which lead to a confinement of the solute wave function, are automatically introduced in all the employed quantum-embedding approaches. The largest errors with respect to the experimental data are reported for the CC2-in-HF$_{\text{4w}}$/FQ
and for the MLCC2$_{\text{30CNTOs}}$/FQ 
approaches, for the $n-\pi^*$ and the $\pi-\pi^*$ transitions, respectively. However, for the lowest dark transition, the CC2-in-HF discrepancy is of about 0.06 eV, and when the computed errors are taken into account, it is reduced to about 0.04 eV. This means that all the methods are suitable to accurately reproduce the experimental vacuo-to-water solvatochromic shift of the $n-\pi^*$ excitation. For the second excitation, the largest error is of about 0.12$\pm$0.03 eV for MLCC2$_{\text{30CNTOs}}$/FQ, whereas a perfect agreement is reported for CC2-in-MLHF-AB$_{\text{3w}}$/FQ. 
However, we note that the calculated solvatochromic shifts depend on the value of the excitation energy in the gas-phase, which is highly sensible to the geometry optimization level.\cite{aidas2008performance} As an example, if the gas-phase excitation energies are computed on the same snapshots used for the calculations in aqueous solution by removing all water molecules, then the agreement with the experiments is improved in all cases (see Table S4 given in ESI).

As a final comment on ACRO, we note that in Ref. \citenum{aidas2008performance}, a good prediction of the experimental solvatochromic shift is only achieved by including 12 water molecules in the QM region in a TD-DFT/MM scheme.\cite{aidas2008performance} In this work, we show that an excellent agreement with the experiment can be achieved by only including in the active part three --- in CC2-in-MLHF(-AB)/FQ and MLCC2/FQ ---
or four --- in CC2-in-HF/FQ---
water molecules. Therefore, we can state that the confinement effects introduced by the inactive water molecules are crucial to achieve the correct result.

\subsection{PY in aqueous solution}

\begin{table}[htbp!]
\centering
\begin{tabular}{lcc}
\hline
          & \multicolumn{2}{c}{$n - \pi^*$} \\
Method    & Exc. En. (eV) & Shift (eV) \\
\hline
vacuo                       &  5.01 & -- \\
\hline
exp (vac)                   &  4.59$^a$, 4.63$^b$,  4.74$^c$ & -- \\
\hline
\hline
QM/TIP3P                     & 5.42 $\pm$ 0.02 & -0.41 $\pm$ 0.02 \\
QM/FQ                        & 5.43 $\pm$ 0.02 & -0.42 $\pm$ 0.02 \\
CC2-in-MLHF$_{\text{3w}}$/FQ     & 5.42 $\pm$ 0.02 & -0.41 $\pm$ 0.02 \\
CC2-in-MLHF-AB$_{\text{3w}}$/FQ  & 5.41 $\pm$ 0.02 & -0.40 $\pm$ 0.02 \\
CC2-in-HF$_{\text{4w}}$/FQ & 5.39 $\pm$ 0.02 & -0.38 $\pm$ 0.02 \\
MLCC2$_{\text{36CNTOs}}$/FQ    & 5.37 $\pm$ 0.02 & -0.36 $\pm$ 0.02 \\
\hline 
exp (wat)                    &  4.94 & -0.31$\pm$0.04 \\
\hline
\end{tabular}
\caption{$n-\pi^*$ excitation energies and vacuo-to-solvent solvatochromic shifts of PY in vacuo and in aqueous solution. Computed standard errors at the 68\% confidence interval are also given. The experimental excitation energy of PY in aqueous solution is taken from Ref. \citenum{marenich2014electronic}. \\
$^a$ Reproduced from Ref. \citenum{schreiber2008benchmarks} \\
$^b$ Reproduced from Ref. \citenum{marenich2011practical}\\
$^c$ Reproduced from Ref. \citenum{cai2000low}\\}
\label{tab:py_w}
\end{table}

In the case of pyridine (PY), we focus on the dark $n - \pi^*$ transition, for which a vacuo-to-water solvatochromic shift of about 0.3-0.4 eV is reported in the literature.\cite{marenich2014electronic} 
We calculate the gas-phase excitation energy at the geometry optimized using B3LYP/aug-cc-pVDZ level. We obtain an absorption energy of 5.01 eV at CC2/aug-cc-pVDZ level, which is shifted from the experimental data of about 0.3-0.4 eV (see Tab. \ref{tab:py_w}), similarly to the case of ACRO. However, we note that different experimental gas-phase data are reported in the literature, ranging from 4.59 to 4.74 eV.\cite{schreiber2008benchmarks,marenich2011practical,cai2000low} Among them, the 4.63 eV value is taken as reference for the calculation of the vacuo-to-water solvatochromic shift,\cite{marenich2014electronic} and we assign an uncertainty of about 0.04 eV. 

\begin{figure}[htbp!]
\centering
\includegraphics[width=.4\textwidth]{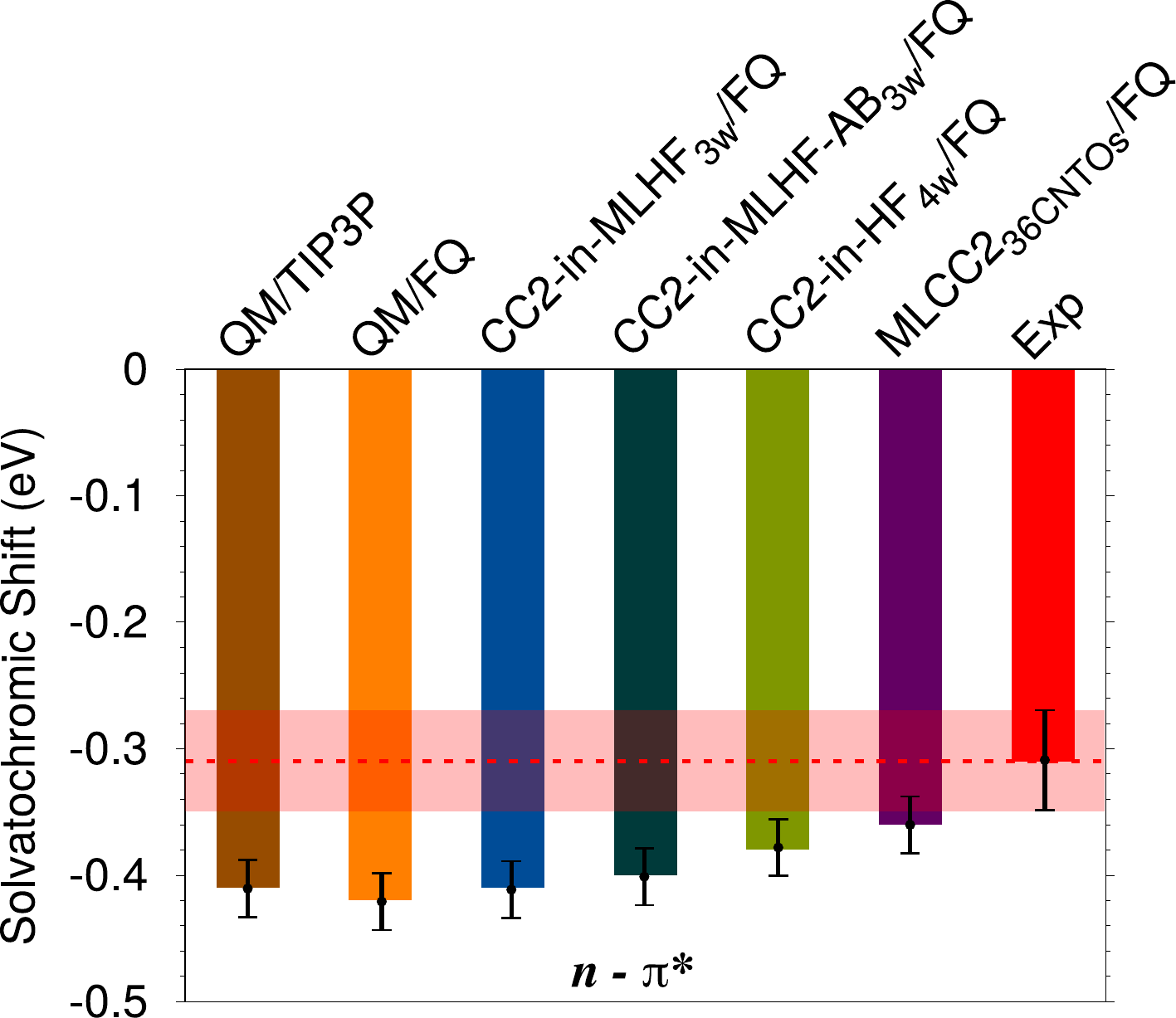}
\caption{Computed and experimental $n - \pi^*$ vacuo-to-water solvatochromic shifts of PY (Values are given in Tab. \ref{tab:py_w}). Computed and experimental standard error bars are also plotted. The horizontal dashed line and the red box follow the experimental value and its error bar.}
\label{fig:PY_MD}
\end{figure}

The absorption energies of PY in aqueous solution are computed by means of QM/TIP3P, polarizable QM/FQ, CC2-in-MLHF(-AB)/FQ, CC2-in-HF/FQ and MLCC2/FQ approaches. In particular, three and four water molecules are included in the active part of CC2-in-MLHF(-AB) and CC2-in-HF
calculations, respectively. In MLCC2 we considered 36 active CNTOs, which correspond to PY and three water molecules. The computed absorption energies and the vacuo-to-water solvatochromic shifts are reported in Tab. \ref{tab:py_w} and in Fig. \ref{fig:PY_MD}, 
along with their experimental counterparts. The experimental excitation energy in aqueous solution is taken from Ref. \citenum{marenich2014electronic}. 

The excitation energies computed using the different approaches range from 5.37 eV with MLCC2$_\text{36CNTOs}$/FQ to 5.43 eV with QM/FQ, 
and are shifted of about 0.4 eV with respect to the experimental value. The corresponding solvatochromic shift ranges from -0.36 eV to -0.42 eV. Also for the reproduction of solvatochromism, the QM/FQ 
method yields the worst agreement with the experiment, whereas the best agreement is shown for MLCC2$_\text{36CNTOs}$/FQ. This confirms the results already commented for ACRO $n-\pi^*$ excitation, which presents the same trends. 

From the investigation of Fig. \ref{fig:PY_MD}, we note that only MLCC2$_{\text{36CNTOs}}$/FQ accurately reproduce the experimental value (within its uncertainty range). However, all the other quantum-embedding/FQ methods present an absolute error ranging from 0.01 (CC2-in-HF$_{\text{4w}}$/FQ) to 0.04 eV (CC2-in-MLHF$_{\text{3w}}$/FQ), thus resulting in an almost perfect agreement with the experimental reference.

\subsection{PNA in aqueous solution}

\begin{table}[htbp!]
\centering
\begin{tabular}{lcc}
\hline
          & \multicolumn{2}{c}{$\pi - \pi^*$} \\
Method    & Exc. En. (eV) & Shift (eV) \\
\hline
vacuo                          &  4.38 & -- \\
\hline
exp (vac)                      &  4.25 & -- \\
\hline
\hline
QM/TIP3P                       &  3.77 $\pm$ 0.01 & 0.61 $\pm$ 0.01 \\
QM/FQ                          &  3.43 $\pm$ 0.01 & 0.95 $\pm$ 0.01 \\
CC2-in-MLHF$_{\text{3w}}$/FQ     &  3.61 $\pm$ 0.01 & 0.77 $\pm$ 0.01 \\
CC2-in-MLHF-AB$_{\text{3w}}$/FQ  &  3.66 $\pm$ 0.01 & 0.72 $\pm$ 0.01 \\
CC2-in-HF$_{\text{4w}}$/FQ &  3.54 $\pm$ 0.01 & 0.84 $\pm$ 0.01 \\
MLCC2$_{51\text{CNTOs}}$/FQ            &  3.47 $\pm$ 0.01 & 0.93 $\pm$ 0.01 \\
\hline 
exp (wat)                      &  3.26  & 0.99  \\
\hline
\end{tabular}
\caption{$\pi-\pi^*$ excitation energies and vacuo-to-solvent solvachromic shifts of PNA in vacuo and in aqueous solution. Computed standard errors at the 68\% confidence interval are also given. Vacuo and water experimental values are reproduced from Ref. \citenum{millefiori1977electronic} and Ref. \citenum{kovalenko2000femtosecond}, respectively.}
\label{tab:pna_w}
\end{table}

\begin{figure}[htbp!]
\centering
\includegraphics[width=.4\textwidth]{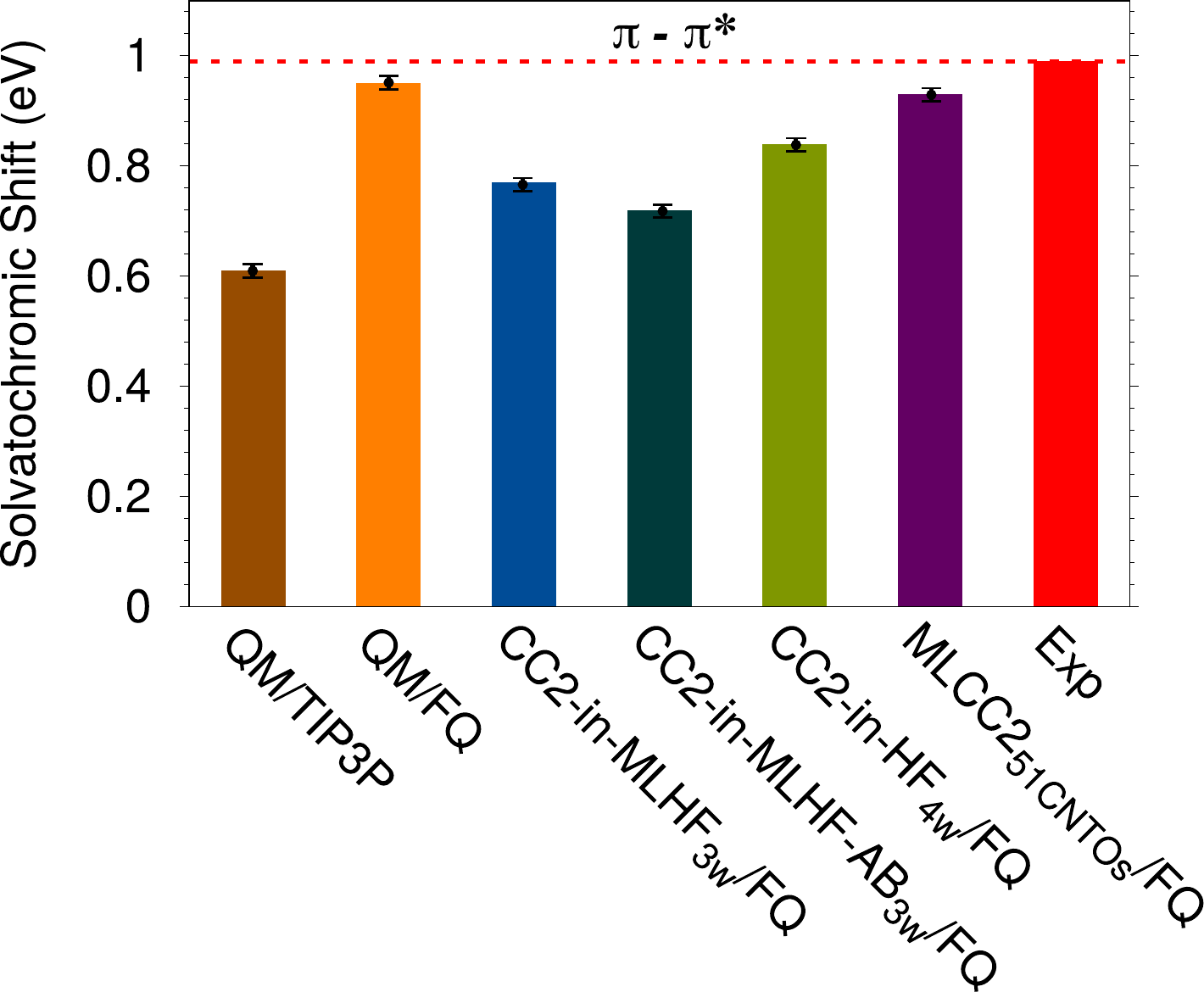}
\caption{Computed and experimental $\pi - \pi^*$ vacuo-to-water solvatochromic shifts of PNA (Values are given in Tab. \ref{tab:pna_w}). Computed standard error bars are also plotted. The horizontal dashed line follows the experimental value.}
\label{fig:pna_MD}
\end{figure}

Finally, we investigate the vacuo-to-water solvatochromic shift of the bright $\pi - \pi^*$ transition of para-nitroaniline (PNA). The solvatochromism of the PNA moiety has been theoretically studied in previous works.\cite{marenich2014electronic,duchemin2018bethe,kosenkov2010solvent,giovannini2019fqfmulinear,frutos2013theoretical} From the experimental point of view, a solvatochromic shift of about 0.99 eV has been measured.\cite{kovalenko2000femtosecond,millefiori1977electronic,marenich2014electronic} As reported in Ref. \citenum{giovannini2019fqfmulinear}, similarly to the other molecules, the gas phase excitation energy is computed at the CC2/aug-cc-pVDZ level using the CAM-B3LYP/aug-cc-pVDZ optimized geometry. Both vacuo and aqueous computed absorption energies are reported in Tab. \ref{tab:pna_w}, together with their experimental counterparts. The vacuo-to-water solvatochromic shifts are also given, and graphically depicted in Fig. \ref{fig:pna_MD}. The MLCC2 calculations are performed by including 51 CNTOs in the active part, which correspond to PNA and three water molecules.

In this case, the computed and experimental gas phase values are in a good agreement, with a discrepancy of 0.13 eV. The vacuo-to-water solvatochromic shifts computed using the different methods employed to model solvation range from 0.61 eV with QM/TIP3P to 0.95 with QM/FQ.
All the results are in line with what has already been commented for ACRO $\pi-\pi^*$ transition. In fact, the excitation energy in aqueous solution decreases with respect to the gas phase, resulting in a positive solvatochromism. Also, the two QM/MM methods predict the worst (QM/TIP3P) and the best (QM/FQ) agreement with the experiment. However, similarly to ACRO, such a result is due to error cancellation, in particular the neglection of any solute-solvent Pauli repulsion.

We now move on to comment on the results obtained using the investigated quantum-embedding approaches. Among them, the largest discrepancy with the experiment is reported for both CC2-in-MLHF/FQ and CC2-in-MLHF-AB/FQ. A slight improvement is shown by CC2-in-HF/FQ,
for which the computed error is about 0.16 eV. We recall that for CC2-in-HF the reference state is represented by a HF calculation on the full QM system, whereas in CC2-in-MLHF(-AB) the reference state is obtained by partitioning the QM system into the active and inactive parts. As a consequence, for CC2-in-HF the mutual polarization between the active and the inactive components is fully taken into account, whereas in CC2-in-MLHF(-AB) it is only included at the first step of the computational protocol, i.e. diagonalizing the starting Fock matrix. We can therefore ascribe the better performance of CC2-in-HF to the importance of polarization terms in the reference state. Finally, among the studied quantum-embedding approaches, the most accurate result is observed for MLCC2$_{\text{51CNTOs}}$/FQ, for which the error with respect to the experiment is of about 0.06 eV. We note that the better performance of MLCC2 can be related to the inclusion of polarization and dispersion terms at the CC level. 

\section{Summary and Conclusions}

In this work, we have presented the coupling of different quantum-embedding approaches with a third MM layer, which can be either polarizable or non-polarizable. In particular, such coupling is discussed for the multilevel families of methods, such as CC2-in-MLHF, CC2-in-MLHF-AB, CC2-in-HF and MLCC2, which have been recently developed in our group.
The resulting three layers approaches are particularly useful, because the computational cost of the quantum-embedding methods can be reduced by treating the long-range interactions at the MM level.

We have applied the developed methods to the calculation of the solvatochromic shifts of molecular systems in aqueous solution, for which an atomistic description of the environment is crucial to correctly describe the specific solute-solvent interactions, such as hydrogen bonding. The absorption energies have been calculated at the CC2 level of theory, which guarantees a good compromise between the computational cost and the accuracy. In particular, we have selected three moieties, ACRO, PY and PNA, for which experimental 
vacuo-to-solvent solvatochromic shifts have been reported in the literature. 

Firstly, we have discussed a benchmark of the methods for a randomly selected snapshot of ACRO, for which we have shown that the third MM layer can accurately take into account the shift related to the long-range interactions. As a consequence, the quantum region can be limited to the first solvation shell. We have shown this to be sufficient to take into account the most relevant short-range interactions, such as Pauli repulsion. Among the different investigated approaches, we show that the MLCC2/FQ method can provide an accurate prediction of both $n-\pi^*$ and $\pi-\pi^*$ excitations, which are the most diffuse transitions in organic molecules. This can be achieved by only including a number of CNTOs corresponding to the solute and three water molecules. A good agreement with the experiment is achieved by including three or four waters in all the other approaches, although the accuracy of the results strongly depends on the considered system. 

To conclude, we point out that the polarization in polarizable QM/MM calculations is only included in the ground state reference. However, it can also be introduced in the following CC calculations.\cite{caricato2014corrected,caricato2010coupled,caricato2010electronic,caricato2013implementation,caricato2018coupled,cammi2012coupled} Such an extension, together with the inclusion of QM/MM Pauli repulsion\cite{giovannini2017disrep} at the CC level, will be the topic of future communications. Finally, we note that the three layer approach that we have developed can also be applied to any QM/classical approach, as for instance the polarizable continuum model (PCM),\cite{tomasi2005,Mennucci12_386} due to its theoretical similarity with the QM/FQ approach.\cite{giovannini2019tpa}

\section*{Conflicts of interest}

There are no conflicts to declare.

\section*{Acknowledgements}

We acknowledge funding from the Marie Sklodowska-Curie European Training Network “COSINE - COmputational Spectroscopy In Natural sciences and Engineering”, Grant Agreement No.  765739, and the Research Council of Norway through FRINATEK projects 263110 and 275506.


\balance


\bibliography{biblio} 
\bibliographystyle{rsc} 

\end{document}